\documentclass[preprint,journal]{vgtc}                

\PassOptionsToPackage{dvipsnames}{xcolor}
\vgtcinsertpkg

\usepackage{microtype}                 
\PassOptionsToPackage{warn}{textcomp}  
\usepackage{textcomp}                  
\usepackage{times}                     
\usepackage{booktabs} 
\usepackage[T1]{fontenc}
\usepackage[utf8]{inputenc}

\usepackage{cite}  

\usepackage{adjustbox}
\usepackage{xspace}
\usepackage{relsize}
\def\code#1{{{\relsize{-1}\texttt{#1}}}\xspace}
\usepackage{amsmath}
\usepackage{listings}
\usepackage{amssymb}
\usepackage[labelfont=sf]{subcaption}
\usepackage{cleveref}

\if11
\def\iw#1{\textsc{\color{green}IW: #1}}
\def\sz#1{\textsc{\color{orange}SZ: #1}}
\def\will#1{\textsc{\color{blue}Will: #1}}
\def\nvm#1{\textsc{\color{red}Nate: #1}}
\else
\def\iw#1{}
\def\sz#1{}
\def\will#1{}
\def\nvm#1{}
\fi

\def\gear{\emph{NASA Landing Gear}\xspace}
\def\exajet{\emph{Exajet}\xspace}
\def\cloud{\emph{TAC Molecular Cloud}\xspace}
\def\impact{\emph{LANL Impact}\xspace}
\def\stellar{\emph{Princeton Stellar Cluster Wind}\xspace}

\newcommand{\Test}[1]{\expandafter\hat#1}

\onlineid{1059}

\vgtccategory{Research}
\vgtcpapertype{algorithm/technique}

\title{Ray Tracing Structured AMR Data Using ExaBricks\vspace{-0.5em}}

\author{Ingo Wald, Stefan Zellmann, Will Usher, Nate Morrical, Ulrich
Lang, and Valerio Pascucci}

\authorfooter{
\item Ingo Wald is with NVIDIA, iwald@nvidia.com,
\item Stefan Zellmann and Ulrich Lang are with the University of
    Cologne, Chair of Computer Science.
\item Will Usher is with the SCI Institute, University of Utah and
    Intel Corp.
\item Nate Morrical and Valerio Pascucci are with the SCI
    Institute, University of Utah.
}

\teaser{
  \vspace{-0.75em}
  \centering \resizebox{0.98\textwidth}{!}{
  }
  \resizebox{0.98\textwidth}{!}{
    \includegraphics[height=6cm]{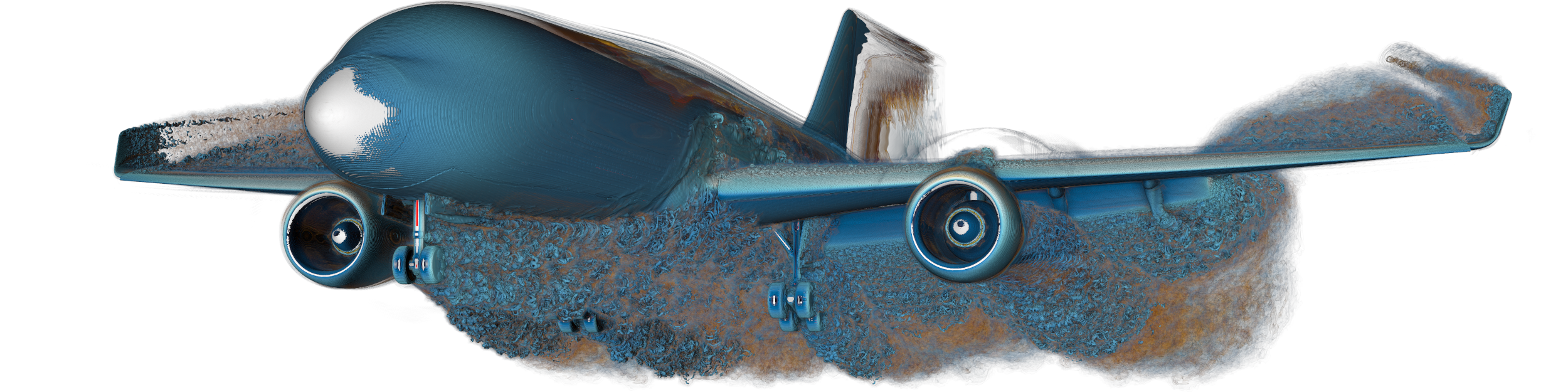}
  }
  \vspace{-0.5em}
  \caption{The \exajet contains an AMR simulation of air flow
    around the left side of a plane, and consists of 656M 
    cells (across four refinement levels)
    plus 63.2M triangles. For rendering we mirror the data set via
    instancing, resulting in effectively 1.31B instanced cells and 126M
    instanced triangles. This visualization---rendered
    with our method---shows flow vorticity and velocity, with an
    implicitly ray-traced iso-surface of the vorticity (color-mapped
    by velocity), plus volume ray tracing of the vorticity field. At a
    resolution of 2500$\times$625, and running on a workstation with
    two RTX~8000 GPUs, this configuration renders in roughly 252
    milliseconds.
    \label{fig:teaser}
    \vspace{-0.5em}
} }
\abstract{Structured Adaptive Mesh Refinement (Structured AMR)
  enables simulations to adapt the domain resolution to save
  computation and storage, and has become one of the dominant data
  representations used by scientific simulations; however, efficiently
  rendering such data remains a challenge. We present an efficient
  approach for volume- and iso-surface ray tracing of Structured AMR data
  on GPU-equipped workstations, using a combination of two
  different data structures. Together, these data structures allow a ray tracing
  based renderer to quickly determine which segments along the ray
  need to be integrated and at what frequency, while also providing quick
  access to all data values required for a smooth sample
  reconstruction kernel. Our method makes use of the RTX ray tracing
  hardware for surface rendering, ray marching, space skipping, and
  adaptive sampling; and allows for interactive changes to the
  transfer function and implicit iso-surfacing thresholds. We
  demonstrate that our method achieves high performance with little
  memory overhead, enabling interactive high quality rendering of
  complex AMR data sets on individual GPU workstations.}


\definecolor{dkgreen}{rgb}{0,0.6,0}
\definecolor{dkblue}{rgb}{0,0,0.6}
\definecolor{gray}{rgb}{0.5,0.5,0.5}
\definecolor{mauve}{rgb}{0.58,0,0.82}

\definecolor{commentgreen}{RGB}{2,112,10}
\definecolor{eminence}{RGB}{108,48,130}
\definecolor{weborange}{RGB}{255,165,0}
\definecolor{frenchplum}{RGB}{129,20,83}

\begin{document}
\maketitle


\firstsection{Introduction}

\section{Introduction}

In many large-scale simulations performed today, the features being simulated 
are quite small relative to the computational domain. For example, the turbulent vortices formed through 
airflow over an airplane can be centimeters in size, but are the product of 
complex interactions between the air and geometry over the scale of the entire 
system (\Cref{fig:teaser}). Other examples can be found in astrophysics,
where scientists are interested 
in planetary-scale forces interacting over light-years of space; or in 
engineering, where millimeter-scale combustion effects are simulated in the 
context of a 20-story boiler.

To account for these massive spatial differences, modern simulation codes 
employ Adaptive Mesh Refinement (AMR)~\cite{chombo,
  grchombo,Fryxell_2000,lava,xrage,germain2000uintah,enzo
}: As the simulation progresses, 
an initially coarse grid is adaptively refined to preserve fine details. The 
output of such simulations are data sets containing significant 
differences in spatial resolution across the computational domain. For example, 
the ratio of largest to smallest cell size in the \gear data set
is 4096 to 1 (\Cref{fig:models}).

Although AMR has become increasingly common in simulations,
visualizing the resulting data continues to pose a number of challenges.
First, different simulation codes use different techniques for 
refinement, resulting in a number of different
forms of AMR that a visualization pipeline must 
support. Moreover, for cell-centered AMR, 
artifact-free rendering requires a method of reconstructing a continuous 
scalar field from the discrete samples.
As AMR data is inherently irregular, computing such samples
requires non-trivial (and costly) operations such as tree traversals to query cell
values.

Especially when reconstructing this scalar field across boundaries between 
different resolution
levels, such methods are neither trivial nor cheap.
As a result, current visualization approaches
either significantly reduce the number of samples taken to achieve
interactive rendering, sacrificing quality, or must take
many expensive samples to achieve high-quality, sacrificing
performance.


In this paper, we present a holistic approach to address the above
issues. The core idea of our approach is to use a combination of three
different but inter-operating data structures that address different
parts of the problem: First, we avoid looking at individual cells,
deep AMR hierarchies, octrees, etc; and instead re-arrange the data
into a set of \emph{bricks}, similar to previous approaches that
reorganized the data or built additional
hierarchies~\cite{kaehler:amr,wald:17:AMR,wang:20:tamr}.
Second, on top of these bricks we build
an \emph{additional} spatial partitioning that is particularly
designed for the AMR \emph{basis} reconstruction
filter~\cite{wald:17:AMR} that
stores for each region which bricks can influence the
region. The regions can then be used for fast \emph{basis}
reconstruction during rendering without costly cell location kernels,
along with space skipping and adaptive sampling. Third, we build an RTX
BVH over the resulting regions, and use this for hardware-accelerated
ray marching, space skipping, and adaptive sampling, while also
supporting interactive transfer function and iso-surface editing.



Our approach supports interactive high-quality direct volume rendering
with a smooth AMR reconstruction filter and gradient-based volume
shading, combined with surface shading from implicit iso-surfaces
and/or polygonal surfaces defined throughout the volume.
In particular, our adaptive sampling approach guarantees that the sampling
frequency can adapt to the finest level cell size 
while keeping total sample counts tractable.
When combined with our fast
sample reconstruction and RTX-accelerated ray marching, this allows for
interactive visualization on individual GPU workstations, even
for highly complex models.
Our key contributions are:
\begin{itemize}
    \setlength\itemsep{-0.1em}
    \setlength\topsep{0em}
    \item
A data structure to reorganize AMR data that supports high-quality
cell interpolation, including across level boundaries;
    \item
A novel method to compute gradient vectors from just the samples
taken for reconstruction;
    \item
An adaptive sampling and opacity correction method for adaptive
sampling of AMR data; and
    \item
A thorough evaluation on realistic models and rendering methods.
\end{itemize}
\section{Related Work}

AMR data sets topologically resemble hierarchies of
structured grids that store scalar fields, lending
themselves to the typical rendering modalities for this type of data, namely,
direct volume or iso-surface rendering.
For the latter, Weber et
al.~\cite{weber2003extraction,weber2012efficient} proposed a scheme
that extracts crack-free iso-surfaces from AMR data by first computing the dual mesh
using unstructured mesh elements, and then extracting an iso-surface from
the unstructured mesh~\cite{weber2003extraction}.
Moran and Ellsworth~\cite{moran:dual} improved upon
Weber et al.'s approach by constructing a more general dual-mesh that also works for data
sets in which bricks at level boundaries can differ by more than one
level.

\begin{figure*}
  \centering
    \vspace{-0.5em}
  \begin{subfigure}{0.19\linewidth}
    \centering\includegraphics[height=9.0em,trim={2cm 3cm 0cm 0cm},clip]{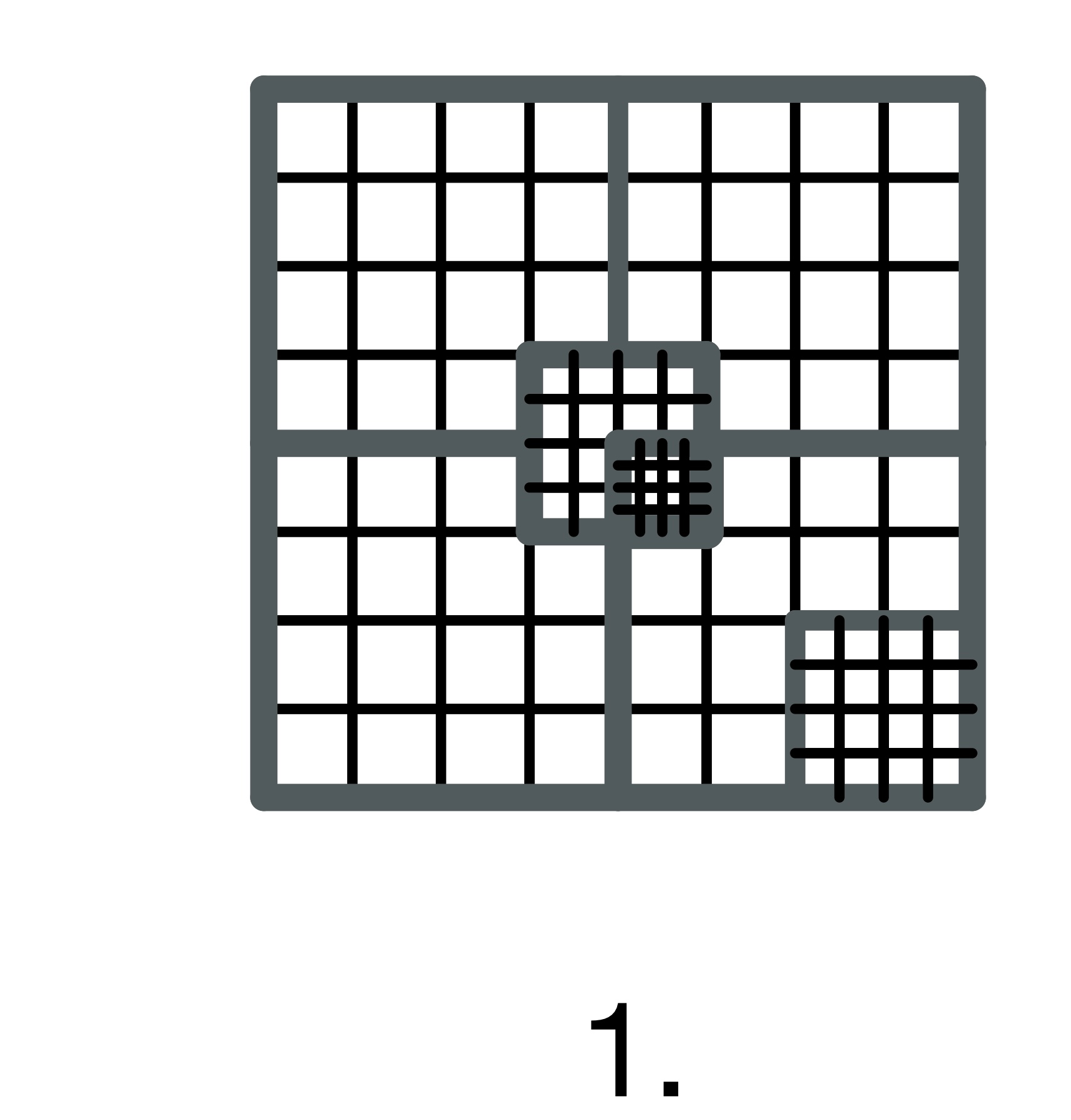}
	\vspace{-1.25em}
    \caption{\label{fig:fig2a}}
  \end{subfigure}%
  \begin{subfigure}{0.19\linewidth}
    \centering\includegraphics[height=9.0em,trim={2cm 3cm 0cm 0cm},clip]{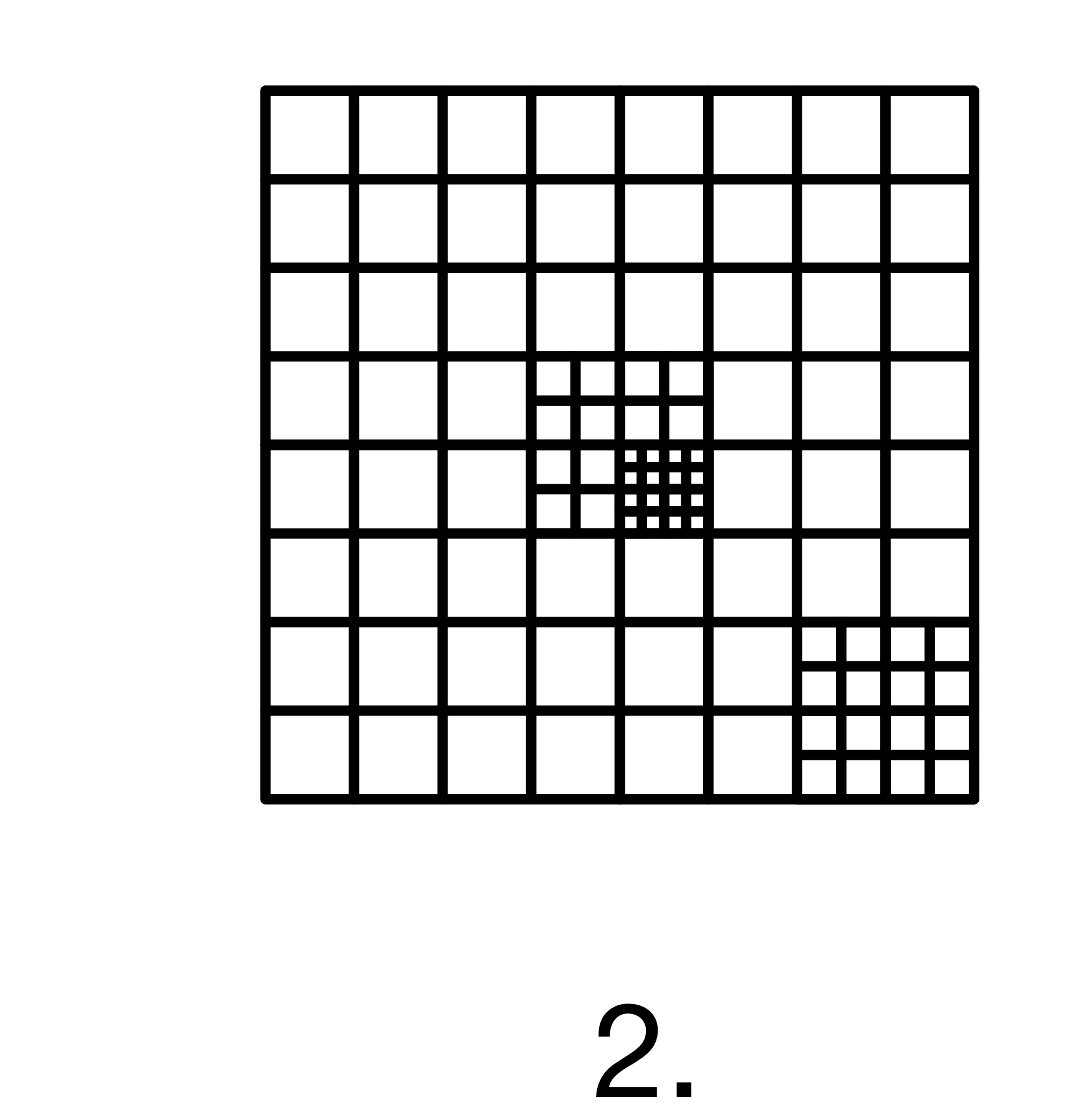}
	\vspace{-1.25em}
    \caption{\label{fig:fig2b}}
  \end{subfigure}
  \begin{subfigure}{0.19\linewidth}
    \centering\includegraphics[height=9.0em,trim={2cm 3cm 0cm 0cm},clip]{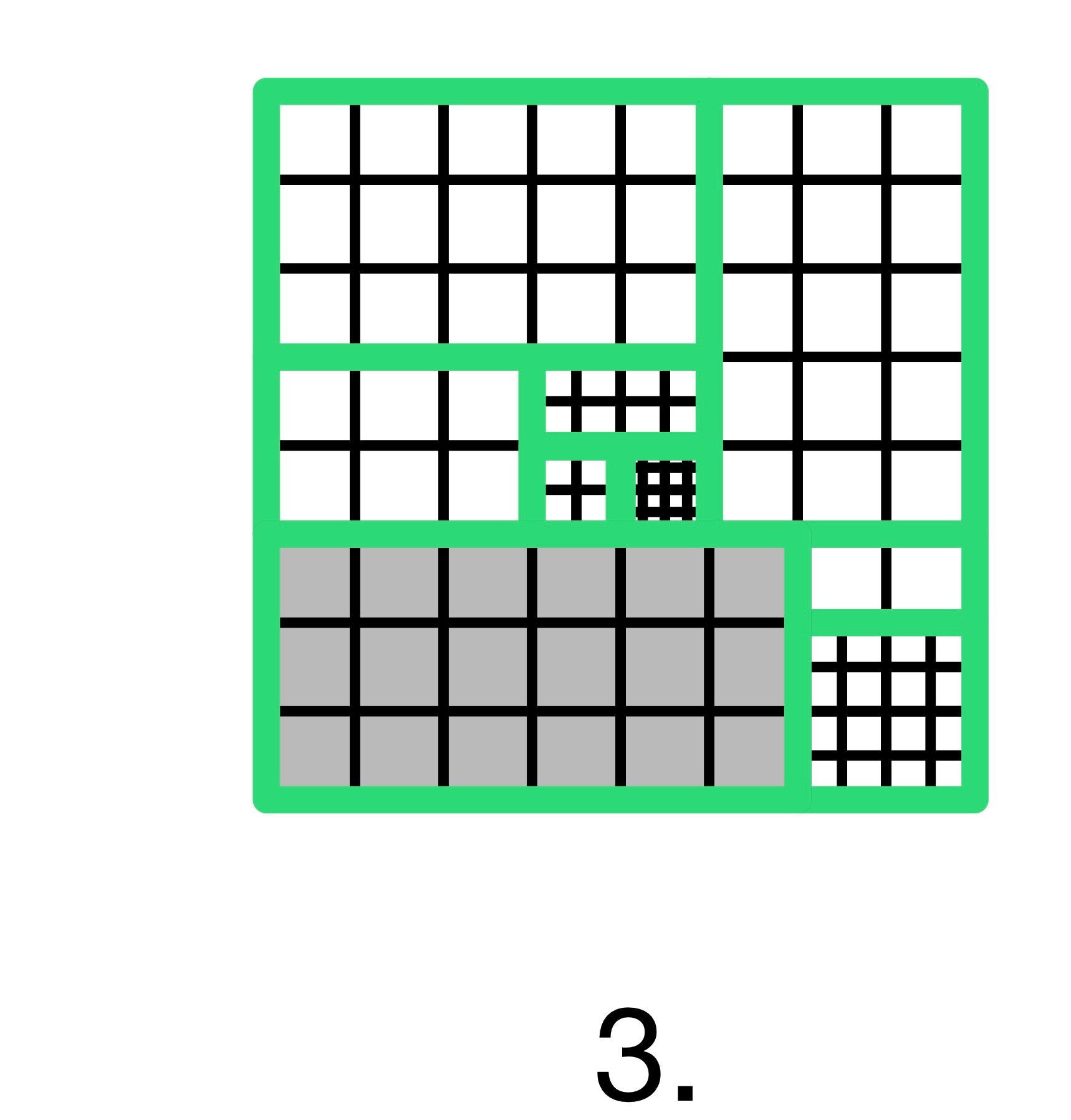}
	\vspace{-1.25em}
    \caption{\label{fig:fig2c}}
  \end{subfigure}
  \begin{subfigure}{0.19\linewidth}
    \centering\includegraphics[height=9.0em,trim={0cm 3cm 0cm 0cm},clip]{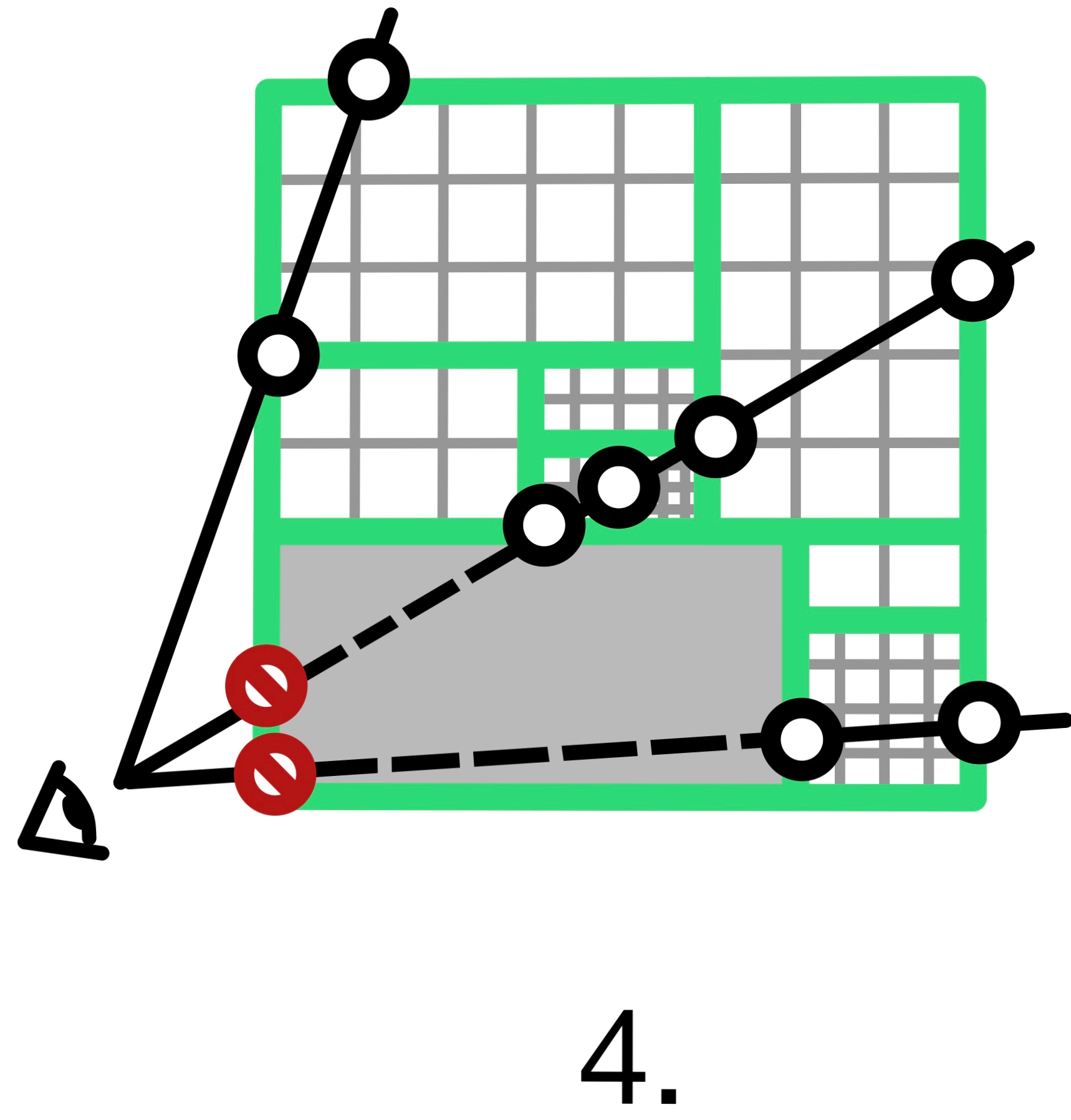}
	\vspace{-2em}
    \caption{\label{fig:fig2d}}
  \end{subfigure}
  \begin{subfigure}{0.19\linewidth}
    \centering\includegraphics[height=9.0em,trim={0cm 3cm 0cm 0cm},clip]{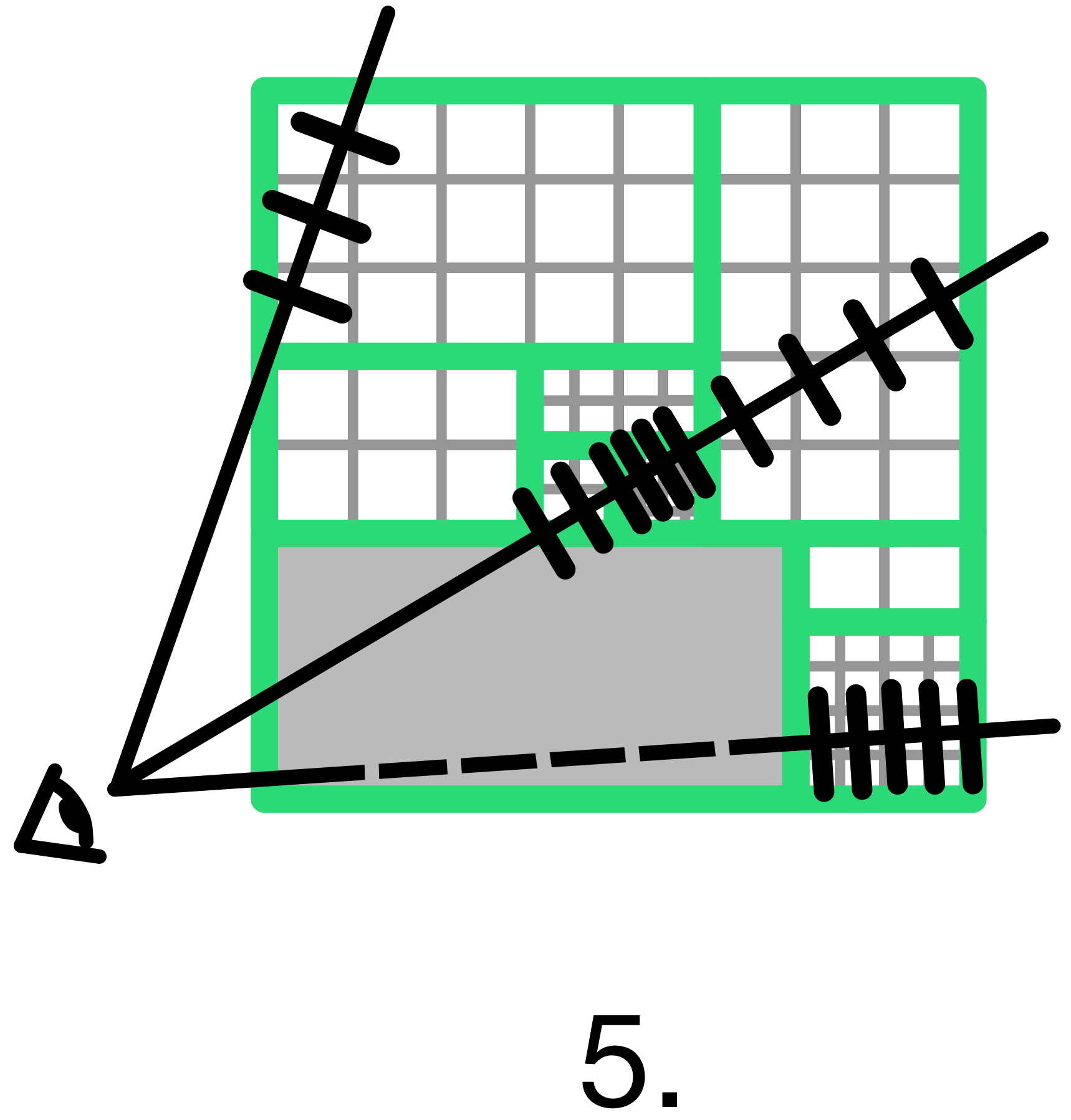}
	\vspace{-2em}
    \caption{\label{fig:fig2e}}
  \end{subfigure}
  \vspace{-1em}
  \caption{\label{fig:method}%
    An illustration of our method
    for the simpler case of a nearest-neighbor reconstruction
    filter: (a) Given an input AMR data set, (b) we first extract the
    finest level cells in each region, discarding the original
    hierarchy. (c) We re-organize the cells into bricks containing
    grids of same-level cells, then for each of the green regions
    compute meta-information for adaptive sampling and space skipping,
    and construct an RTX BVH over these regions. (d) During rendering we
    traverse rays through the bricks, skipping those that are
    transparent. (e) Within each region we adapt the sampling rate to
    the corresponding cell size. 
    When employing a smooth reconstruction filter, 
    both the brick generation (a-c)
    and rendering (d, e) steps are the same;
    however, the Active Brick Regions are traversed instead of the
    bricks.
    \vspace{-1.5em}
  }
\end{figure*}

%

An alternative to extracting iso-surfaces is direct volume
rendering (DVR)~\cite{levoy:1988,drebin:volumeRendering,parker:99:VV,
beyer_state---art_2015}.
Although DVR is widely used for regular structured grids,
in particular on the GPU where hardware texture units can be leveraged for interpolation,
DVR for AMR volumes is far more challenging.
Weber et al.~\cite{weber2003extraction} proposed a method that first
generated unstructured elements to stitch across level boundaries and
then performed a scan conversion on these elements using the GPU.
Park et al.~\cite{park:AMR} proposed a splatting approach, building
off the cell projection method of Max~\cite{max:91:SPC}.
Ma~\cite{ma:cray} also proposed a ray
tracing approach, but assumed that cell-centered AMR could easily be
converted to vertex-centered AMR, which in general is not the case
at level boundaries.

More closely related to our method is the technique proposed by K\"ahler et 
al.~\cite{kaehler:amr}. As in our method, K\"ahler finds a tailored
spatial
partitioning with blocks that contain only same-level cells, discarding
the original AMR grid hierarchy. Their approach uses a multi-pass
rendering method, can leverage 
hardware-accelerated texture sampling on the GPU,
and includes a form of adaptive sampling by adjusting the sampling
rate within a block.
\ifx\empty
Again like our approach, Kähler et al use a ray casting to sample AMR 
bricks and composites samples using front-to-back alpha blending. However, 
for a lack of ray tracing hardware, the authors' partition traversal is limited 
to precomputing raycasting segments by rasterizing the outlines of the block 
bounding boxes, computing per-pixel linked lists in the process which are stored 
in an offscreen render target and then used in a later renderpass.
\iw{far too much detail}
\else
\fi
K\"ahler and
Abel~\cite{kaehler_single-pass_2013} later extended this method
to work in a single rendering pass by using bindless textures and traversing
the $k$-d tree in the fragment shader, though the multipass method
was found to perform best. However, their method only works
for either \emph{nearest} reconstruction or \emph{vertex}-centered
AMR, and does not support smooth reconstruction at all for the cell-centered AMR that
most current simulations use.

\ifx\empty
The simple space decomposition employed by Kähler et al. also enables 
an adaptive sampling approach similar to ours, which requires an 
opacity correction that assumes exponential opacity falloff in a homogeneous 
medium. However, this opacity correction approach when combined with traditional
ray casting leads to artifact at brick boundaries, since the last ray delta used
when sampling through a brick may only be a fraction of that brick's assumed ray 
delta. The authors attempt to avoid these artifacts, although their strategy 
differs from ours (cf. Section~\ref{sec:adaptive}), which we believe to be more 
accurate.
\iw{too defensive, and too detailed}
\fi
%

For rendering Octree AMR data, Labadens et al.~\cite{labadens:octree-amr} proposed using
the octree to generate volume splats, or traced rays through
the octree to perform volume integration at the leaves.
However, their approach also supports only nearest neighbor
interpolation.

Along with iso-surface extraction,
Moran and Ellsworth's~\cite{moran:dual} unstructured dual-mesh
framework is also capable of high-quality volume rendering of AMR
data.
Their framework did not aim for interactive rendering, but
produced high quality imagery and scaled to large models.
To achieve high-quality volume rendering, Moran and Ellsworth introduced
an adaptive sampling approach which adjusts the sampling rate
to match the local data frequency.

To avoid the need to construct an unstructured dual-mesh or
unstructured stitching elements, Wald et al.~\cite{wald:17:AMR}
and Wang et al.~\cite{wang:18:iso-amr}
proposed several reconstruction filters that can operate directly on the
cell-centered AMR data.
As with our presented work, these papers aimed at
interactive rendering of AMR data within a ray tracing framework (in their
case, OSPRay~\cite{ospray}), and support both direct volume and
implicit iso-surface rendering. However, both methods
still suffer from two main shortcomings. First, both require
the frequent use of costly cell location kernels, making
computing samples expensive. Second, they do not
introduce any space skipping or adaptive sampling tailored for AMR data, 
and instead rely on OSPRay's existing adaptive sampling
code, which is not aware of the underlying AMR hierarchy
and thus may severely under- or over-sample the data. 
In this paper, we adopt the \emph{basis} interpolation method
introduced in~\cite{wald:17:AMR} to provide a continuous interpolant
across level boundaries, but do not require the
costly cell location kernels that limited the original method's performance.

A key difference of our approach using the \emph{basis} method
compared to prior AMR rendering work by Leaf et
al.~\cite{leaf_efficient_2013} leveraging the multiresolution
interblock interpolation of Ljung et
al.~\cite{ljung_multiresolution_2006}, is that the basis
function's support is not truncated at the first sample on the
neighboring side. 
Although this leads to a smoother interpolant, it also results in
larger support overlap between cells from neighboring blocks.

Wang et al.~\cite{wang:20:tamr} recently presented a
reconstruction algorithm for high-quality rendering of Tree-Based AMR data,
combined with a sparse octree representation for traversal.
Their interpolant works by virtually introducing unstructured elements
to stitch across boundaries, which fall into a fixed number of cases
for which the interpolation weights can be precomputed. They achieve
empty space skipping and adaptive sampling through their sparse octree.
However, the large number of top-down octree traversals required for
sample reconstruction in their approch impacts rendering performance.

With regard to empty space skipping, Ganter and
Manzke~\cite{ganter:19:optix} (for structured volumes) and Morrical et
al.~\cite{morrical:19:adaptive} (for unstructured volumes) proposed to
leverage modern graphics hardware for on-the-fly space skipping
through hardware-accelerated ray tracing. For unstructured data,
Morrical et al.~\cite{morrical:19:adaptive} further improve rendering
performance by adapting the sampling rate to the data variation within
regions of a spatial subdivision that they computed over the
unstructured elements; they also used RTX capabilities for marching
through these regions. In this work, we adopt a similar strategy
for space skipping and adaptive sampling, but on top of our
proposed data structure.
\section{The ExaBricks Hierarchy}
\label{sec:data-structure}
\label{sec:method}

The core idea of our method is to use a combination of three different
inter-operating data structures and associated algorithms to jointly address
the different aspects of the problem to efficiently render AMR data
with smooth interpolation.

First, we re-organize the input AMR cells into a set of compact,
non-overlapping \emph{bricks} of same-level cells, as proposed by K\"ahler
et al.~\cite{kaehler:amr}. These
bricks can be extremely small, e.g. on the \exajet
bricks with 1, 2, or 4 cells are in fact quite common at level
boundaries; however, 
in larger homogeneous regions these bricks allow for storing cells in
a memory-efficient manner (Section~\ref{sec:bricks}).

Second, for each of these bricks we compute the region of space
where the reconstruction filters associated with the brick's cells
are non-zero. When using a \emph{nearest} reconstruction filter these regions are just the
brick's bounds; however, for any smooth interpolant these
regions extend beyond the brick and result in overlapping regions of support (see
Figure~\ref{fig:regions}).
We compute a second spatial partitioning
structure over these support regions, the Active Brick Regions (ABRs), to produce a data
structure where each leaf stores a list of the bricks
that potentially influence the spatial region covered by
the leaf.
For any smooth interpolant, the resulting ABRs are more complex
in shape and number than the original bricks.
The ABRs are best seen as the ``glue'' of our method, that allow us to
combine the bricks for low overhead storage (Section~\ref{sec:bricks}),
smooth reconstruction filters and cell location without
top-down queries (Section~\ref{sec:regions}),
space skipping (Section~\ref{sec:space-skipping}),
and adaptive sampling (Section~\ref{sec:adaptive}) into a single coherent method.

%

Finally, we build an RTX BVH over the ABRs that we use to quickly
iterate over those that a given ray needs to integrate,
while skipping those it is safe to ignore (Section~\ref{sec:BVH}).

\ifx\empty
In the remainder of the paper, Section~\ref{sec:data-structure}
describes our core \emph{ExaBricks} data structure and how we
build the bricks. Sections~\ref{sec:nearest} and~\ref{sec:basis}
discuss how we then use those bricks for fast sample reconstruction, space
skipping, and adaptive sampling; first introducing these concepts for
the simpler case of nearest-neighbor reconstruction in
Section~\ref{sec:nearest}, then extending these same concepts to
basis method reconstruction in
Section~\ref{sec:basis}. Section~\ref{sec:implementation} then
discusses our particular implementation of this method within a
non-trivial rendering framework, followed by results in
Section~\ref{sec:results}. We discuss pros and cons of this approach
and conclude in Section~\ref{sec:discussion}.

On a conceptual level we use the same two core ingredients as Wald
et al.~\cite{wald:17:AMR}---namely, a ray tracer that combines ray
traced surface-rendering with ray marching based direct volume
rendering; and the \emph{basis method} reconstruction filter for
interpolating the AMR data. However, we use a combination of techniques that
together significantly reduce the volume sampling cost which 
dominated rendering time in previous work.

Our approach is built on a novel data structure--which we call
\emph{ExaBricks}---that works as follows (\Cref{fig:method}):
\begin{enumerate}
\item
  We re-organize the input AMR cells into multiple 3D grids of
  same-level cells, which are built to be as large as possible so as
  to minimize the number of bricks, and consequently, the number of
  subsequent brick traversal operations (\Cref{fig:fig2c}). 
\item We build a spatial partitioning over our model that contains
  either bricks (for nearest-neighbor interpolation

  We build a spatial partitioning over the basis functions'
  \emph{supports} of these bricks, which for any point in space can
  directly tell us which bricks we must interpolate from
  \emph{without} having to re-traverse each sample through a $k$-d tree.
   (\Cref{fig:fig2c}) \nvm{This method overview text seems to be mixed up now 
   with the extension to basis-method interpolation discussed in section 6.
   This method overview should only discuss nearest neighbor interpolation, which 
   doesn't account for basis functions' support regions, as demonstrated in the 
   figure above.}
\item
  We build an acceleration structure over the
  \emph{active brick regions} (i.e., those visible after applying the transfer function)
    allowing us to use
  hardware-accelerated ray tracing to
  quickly iterate through successive visible regions along a
  ray, where each such region will typically yield multiple volume
  samples (\Cref{fig:fig2d}).
\item
  Within each region we automatically adapt the sampling rate based
  on the data frequency of that region's bricks, thereby adapting
  the sampling rate to the data rate (\Cref{fig:fig2e}).
\end{enumerate}

Finally, we integrate our technique into a ray tracer that also
supports ray traced surface-based rendering, implicit
iso-surfacing by root finding, and gradient based shading,
to provide high-quality interactive visualization of large-scale AMR
data.
\fi

\subsection{Organizing Cells into Bricks}
\label{sec:bricks}

\begin{figure}
    \vspace{-1em}
  \centering{
    \resizebox{0.8\columnwidth}{!}{%
      \includegraphics[trim={0cm 0cm 0cm 0cm},clip,height=6cm]{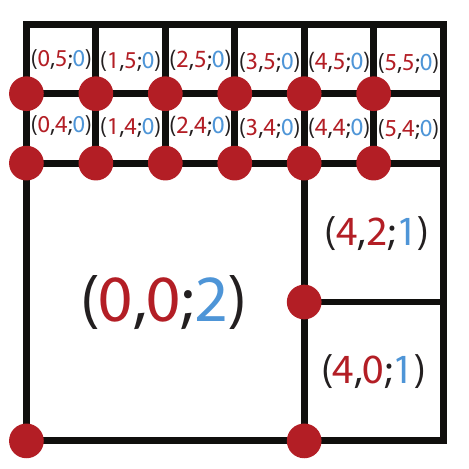}
      \includegraphics[trim={0cm 0cm 0cm 0cm},clip,height=6cm]{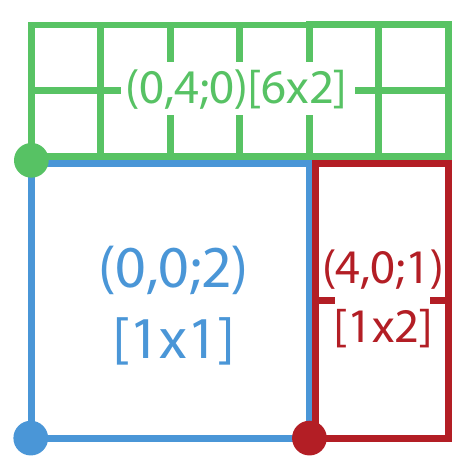}
    }
  }
  \vspace{-1em}
  \caption{\label{fig:bricks-and-cells}%
    Cells and Bricks.
    Left: A 2D illustration of the input AMR cells
    showing 15 cells over three levels, labeled by
    their $(i, j; l)$ coordinates.
    Right: The bricks computed for these input cells,
    labeled by their $(i, j; l)$ coordinates and $[N \times M]$
    dimensions.
    \vspace{-2em}
  }
\end{figure}

The first stage of our method is to (re-)organize
the input data set's AMR cells into a set of
bricks. Our definition and construction of these bricks
is similar to that of K\"ahler et al.~\cite{kaehler:amr},
and thus we focus on introducing the terminology
required for the subsequent sections.


\subsubsection{Cells}

We adopt a terminology where level 0 is the
\emph{finest} level, with an implicit cell size of $1^3$,
level 1 the next coarser level with cells of size $2^3$,
level 2 coarser yet with cells of size $4^3$, and so on.
This follows the terminology used by PowerFlow~\cite{powerflow} and
Cart3D~\cite{cart3d}, but is
the reverse of others such as Chombo~\cite{chombo}.
We can refer to any cell on a given level using four values, $(i,j,k;l)$,
where $(i,j,k)$ are the lower-left unit coordinates of the cell and
$l$ is its level. Each cell contains a single data value.
Cells are arranged by the AMR simulation into blocks of grids
in Block-Structured AMR
layouts~\cite{berger1984adaptive,chombo} or an Octree in
Tree-Based AMR layouts~\cite{cart3d,powerflow}.

Given the refinement rules outlined by Berger and
Oliger~\cite{berger1984adaptive}, we know that $i,j,k$ are all
multiples of $2^l$, and that cells will not overlap. Unlike
Berger and Oliger's constraints, we do not require
cells to be fully refined (``holes'' are explicitly allowed), nor do
we require only single-level differences at level boundaries, making our
method applicable to a wide range of different AMR refinement
schemes. An illustration of the terminology we use is given in
\Cref{fig:bricks-and-cells}.


\subsubsection{Bricks}
Similar to K\"ahler and Abel~\cite{kaehler_single-pass_2013}, we
discard any AMR hierarchy information that is stored in the input
data and reorganize the unordered list of cells into a set of
non-overlapping \emph{bricks} of same-level cells.
The scalars for each brick are stored in a separate array, with a 3D
array of scalars per-brick.
This enables us to
support a wide range of structured AMR data formats (e.g.,
Block-Structured, Octree, Cartesian).
For each brick, we store the coordinates of the lower-left
corner ($i,j,k$), the level ($l$), and the number of cells stored in
each dimension ($N,M,K$) (see~Figure~\ref{fig:bricks-and-cells}).

To generate the bricks, we build a $k$-d tree whose leaf nodes contain only
same-level cells. The tree is built top-down, and a leaf
node created if all cells within the current node are on the same
AMR level and completely fill their combined bounding box.
To avoid bricks becoming so large as to not provide fine enough
granularity for space skipping we limit leaves to be at most 32 cells
wide on any axis.
In contrast to K\"ahler and Abel, the children of an inner node are made
by simply splitting the current cells along the longest axis of the
node's bounding box, which we found to provide better rendering
performance.
Split positions are rounded to an integer multiple of the current
coarsest cell width to ensure cells are never split during the
partitioning.
We then discard the hierarchy and store only the resulting leaves,
corresponding to the bricks.



\nocite{impact-dataset}

\subsection{Basis Method and Active Brick Regions}
\label{sec:regions}

The bricks produced, per the previous section, are readily usable
for rendering with a nearest-neighbor reconstruction filter, as illustrated
in Figure~\ref{fig:method} and described by K\"ahler and
Abel~\cite{kaehler_single-pass_2013}.
However, nearest-neighbor reconstruction
has obvious limitations in terms of image quality, in particular for spiky
transfer functions and/or iso-surfaces. High-quality rendering
requires the use of a more advanced reconstruction filter.

Due to its simplicity and ease of implementation, for this paper we
chose to use the \emph{basis} method by Wald et al.~\cite{wald:17:AMR}
for reconstruction. In this method, each cell
$C$ of width $C_w$ is associated with a hat-shaped basis function:
\begin{equation}
  \label{eq:hat}
  \Test{H_C}(p) = \Test{h}\bigg(\frac{|C_{p_x}-p_x|}{C_w}\bigg)\Test{h}\bigg(\frac{|C_{p_y}-p_y|}{C_w}\bigg)\Test{h}\bigg(\frac{|C_{p_z}-p_z|}{C_w}\bigg),
\end{equation}
where $\Test{h}(x)=\max(1-x,0)$. Reconstructing a sample at position
$p$ then involves finding all cells $C_i$ with data value $C_{v_i}$
that have non-zero support $\Test{H_{C_i}}(p)$ and computing the
weighted sum:
\begin{equation}
  B(p) = \frac{\sum_{C_i}
    \Test{H_{C_i}}(p)C_{v_i}}{\sum_{C_i}\Test{H_{C_i}}(p)}.
  \label{eq:basis}
\end{equation}

\subsubsection{Fast Basis-Method Sample Reconstruction}
\label{sec:fast-basis-reconstruction}

The problem that occurs when using any non-\emph{nearest} reconstruction
filter is that the reconstruction
for a sample may---and typically
will---be influenced by many different cells, potentially from different bricks
at different AMR levels.
To compute the reconstruction for a sample, Wald et al.~\cite{wald:17:AMR}
used a cell location kernel that would execute several recursive $k$-d tree
traversals until all the required cells were found. This method
is elegant, but costly even on a CPU with large caches. To avoid these
per-sample $k$-d tree traversals we instead build a \emph{second} data
structure, the ABRs, that, for each region of space, tells us exactly which
bricks can possibly influence that region. Thus, assuming one is in a
leaf of this data structure, all that is required to perform a
\emph{basis} reconstruction is to iterate over the bricks referenced by
this region, find the cells within each brick
that influence the sample, and add their contributions.
As each brick is a 3D grid, finding the required cells within
the brick is trivial once the brick is known.

\begin{figure}
  \centering
    \vspace{-1em}
  \resizebox{0.98\columnwidth}{!} {
  \begin{subfigure}{0.19\linewidth}
    \includegraphics[height=1.7cm]{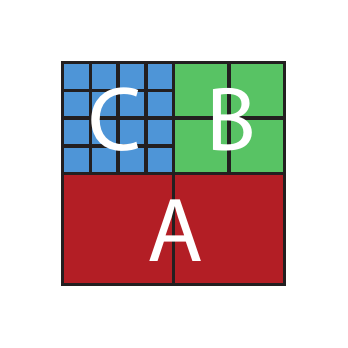}
    \vspace{-4ex}
    \caption{\label{fig:brick-support-a}}
  \end{subfigure}
  \begin{subfigure}{0.19\linewidth}
    \includegraphics[height=1.7cm]{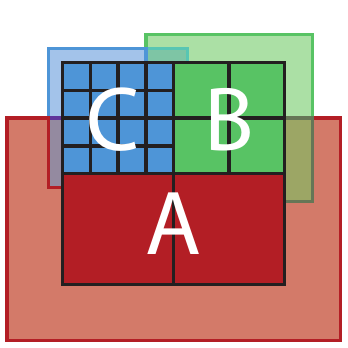}
    \vspace{-4ex}
    \caption{\label{fig:brick-support-b}}
  \end{subfigure}
  \begin{subfigure}{0.19\linewidth}
    \includegraphics[height=1.7cm]{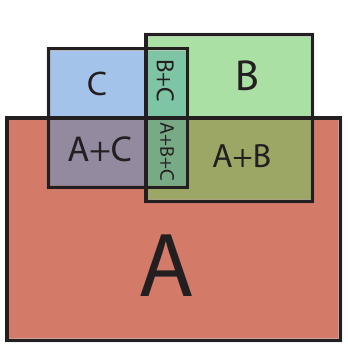}
    \vspace{-4ex}
    \caption{\label{fig:brick-support-c}}
  \end{subfigure}
  \begin{subfigure}{0.19\linewidth}
    \includegraphics[height=1.7cm]{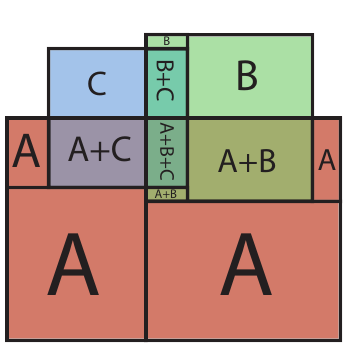}
    \vspace{-4ex}
    \caption{\label{fig:brick-support-d}}
  \end{subfigure}
  }\vspace{-1em}
  \caption{\label{fig:support-regions}\label{fig:regions}
    A 2D Illustration of our Active Brick Regions:
    (a) A data set with three bricks, each of a different
    refinement level.
    (b) The brick support regions corresponding to each brick.
    (c) The overlap of these supports forms a spatial partitioning
    where each region knows which bricks are ``active'' within it.
    (d) We subdivide these regions into non-overlapping rectangular regions which we
    can traverse as before.}
  \vspace{-2em}
\end{figure}

\subsubsection{Extending Bricks to Support Smooth Interpolation}

To compute the ABRs to allow for smooth interpolation across level
boundaries, we define the \emph{support} of a cell $C$ as
the region of space where $\Test{H_C}$ is non-zero. For the
\emph{basis} method, the support is a rectangular box extending
exactly half a cell width beyond what is covered by $C$
itself. Similarly, we define the support of a brick $B$ as the union
of the supports of that brick's cells. Thus, a brick's support is
a box exactly half that brick's cell width larger than the
brick. The exact extent of the brick's support depends on the level
of its cells but, as each brick consists of cells on the same level,
it is always a rectangular region (see~\Cref{fig:support-regions}a and b).

\ifx\empty
Clearly, if at any ray interval $[t_0,t_1]$ we knew exactly which
brick supports we are in,
we could quickly evaluate samples for
any point along the ray interval. One possible method for doing this
would be to store, with each brick,
a list of the bricks whose support overlaps that brick.
However, for very large bricks there can be a large number
of overlapping bricks whose values are not necessarily needed
across the entire brick.

Tracing rays through a BVH built over the brick supports does
not work either: unlike the bricks, the supports overlap,
and tracing a ray through the supports does not guarantee the
successive and non-overlapping ray segments which we require for
integration (see~\Cref{fig:brick-support-b}).
A possible approach to address this is to
use a \emph{multi-hit} kernel~\cite{gribble} to track
which bricks are active within a given ray segment.
However, in an early prototype of this approach we found it difficult to deal with the
fact that we cannot bound \emph{a priori} how many bricks overlap any
given point, making an implementation in a GPU kernel challenging.

We can, however, solve this issue in another way.

\fi

We observe that the superposition of all brick supports forms a
partition of space, where each region can be associated with exactly
those bricks whose supports overlap in that region
(see~\Cref{fig:brick-support-c}). These regions are formed by the
intersection of the various brick supports, and are not necessarily
rectangular or even convex. However, the edges of each region will be
parallel to the coordinate axes, and thus they can be decomposed into
the rectangular non-overlapping Active Brick Regions.
Each ABR tells us exactly which bricks are
``active'' in that region of space, in the sense that at least one
cell from these bricks has non-zero contribution in the region.

For each ABR, we store a list of the brick IDs active in
the region and the bounding box of the region.
We also compute and store additional
meta-information about the region to enable empty space skipping
and adaptive sampling.
To enable empty space skipping, we compute the minimum and maximum
scalar value of any cell whose support overlaps the region.
As the basis method is a linear
combination of the cells' values, this range will
also bound the value of any sample reconstructed in the region.
We also store the cell width of the \emph{finest} level that influences a region to
later adapt the sampling rate for adaptive sampling.



\subsubsection{Constructing the Active Brick Regions}

To construct the active brick regions we again use a recursive
top-down partitioning algorithm, similar to that used for building the
bricks in \Cref{sec:bricks}. We begin by creating a list of
``brick support fragments'', where each fragment specifies a box of
space which it covers and the ID of the brick that is covered by
it. The input list contains one fragment per brick, with each input
fragment covering the brick's entire support region. We then compute
the bounding box of all the fragments, and begin a recursive
subdivision.

In each subdivision step we need to consider a region to subdivide
(initially, the entire bounding box), and a list of brick support
fragments that overlap the region. To find a partitioning plane we
iterate through the set of fragment boundaries and look at their
respective faces, each of which defines a potential partitioning
plane. From these, we select the one that is closest to the current
region's spatial center, preferably along the dimension where the
region is widest. If no such plane exists in any dimension we know
the current region does not contain a support boundary and can create
a leaf with the current set of brick IDs. Otherwise, we partition the
region into two subregions, sort the fragments overlapping the region
into the left and right subtrees, and recursively partition the
subtrees. The output leaves of this partitioning process are the
active brick regions. We show statistics about the number of cells,
bricks, and regions for a number of different AMR data sets
in~\Cref{tab:bricks-and-regions}.


Each active brick region tracks the brick IDs that
may influence the region, allowing us to eliminate the costly top down $k$-d tree
traversals originally required for cell location in
the basis method~\cite{wald:17:AMR}. For each region
we know exactly the bricks influencing it, and can quickly retrieve
the required cells from the brick's 3D array.


\begin{table}
  \vspace{-1em}
  \caption{\label{tab:bricks-and-regions}%
    The number of cells, bricks, and regions for each data set,
    and the average bricks/region, weighted by a simple average
    (by count) and by volume, to account for access probability. }
  \vspace{-1em}
  \centering\relsize{-1}{
  \begin{tabular}{@{}lrrrcc@{}}
      \toprule
      & & & & \multicolumn{2}{c}{Avg. \#Bricks/Region} \\
      \cmidrule(lr){5-6}
      Model & \#Cells & \#Bricks & \#Regions & By Count & By Volume \\
      \midrule
      Cloud & 102M & 528K & 9M & 3.46 & 1.09 \\
      Impact-5K & 26.8M & 515K & 12.3M & 3.42 & 1.24 \\
      Impact-20K & 158M & 3.4M & 89.6M & 3.46 & 1.75 \\
      Impact-46K & 283M & 3.1M & 77.1M & 3.45 & 1.74 \\
      Wind & 411M & 24.6K & 689K & 2.78 & 1.72 \\
      Gear & 262M & 26K & 792K & 2.97 & 1.20 \\
      Exajet & 656M & 3.1M & 55.8M & 3.18 & 1.25 \\
      \bottomrule
      \end{tabular}
  }
  \vspace{-2em}
\end{table}

\subsection{BVH over Active Brick Regions}
\label{sec:BVH}

The active brick regions will allow each
ray to know exactly which bricks influence a certain region of space
and at what frequency; however, we still require a means of
efficiently iterating a ray through the regions it intersects.
One option to do so would be to store the split planes used during
region construction---which form a $k$-d tree---and use this
tree over the regions in the same manner that K\"ahler et
al.~\cite{kaehler:amr} used their $k$-d tree over bricks.
This approach would require implementing
a software $k$-d tree traversal in a shader program with a per-thread stack
and non-trivial control flow, similar to the single-pass approach
of K\"ahler and Abel~\cite{kaehler_single-pass_2013} but over our regions.

We instead leverage the new RTX hardware available on GPUs to
enable hardware-accelerated traversal of the regions.
To do so we discard the $k$-d tree
produced by the region builder and store only the final regions.
We then create an RTX \emph{user geometry} with as many
primitives as we have ABRs to construct an acceleration
hierarchy over them. Our approach for rendering with the produced
BVH is discussed in the following section.


\section{Rendering with the ExaBricks Data Structure}
\label{sec:rendering}




%
%
%
Each ray that is traversed through the RTX BVH in hardware is first
initialized with a search interval $[0..t_\text{max})$. If no region
is found the ray has terminated; otherwise we compute the interval
$[t_\text{in}, t_\text{out}]$ that the ray overlaps with the intersected
region, and volume-integrate this interval over the region as
described below. To find the next ABR we trace another ray starting
from $t_\text{out} + \mathcal{E}$ and repeat until the ray becomes
opaque (for early ray termination), or no subsequent region is found.

Using a new ray traversal for each iteration step means that each
volume ray will perform several hardware ray traversals. However,
thanks to hardware support these rays are cheap, and can typically
be amortized over multiple samples taken within the next region.



\subsection{Space Skipping}
\label{sec:space-skipping}

In addition to amortizing the cost of taking multiple samples
in the region, we also use the ABRs for space skipping.
Truly \emph{empty} space, i.e., areas outside the AMR mesh or holes
in the mesh, will be automatically skipped as such areas do not
generate bricks or regions. The more challenging case for space skipping
are regions that are covered by cells whose visibility depends on the
transfer function. During BVH construction we use each ABR's
precomputed scalar range to compute the maximum opacity of the
transfer function within the range. If the maximum opacity for this
range is 0, we know that every sample taken in this region would
be fully transparent, and consequently that the entire region can be
excluded from the BVH.
Regions that are fully transparent are discarded during BVH
construction by returning an empty box, those that are not
simply return their bounding box.

%
%
%
In particular, we note that we do not have to construct any
\emph{additional} structure for space skipping (e.g., as done by
Morrical et al.~\cite{morrical:19:adaptive}), nor do we have to
check a region's validity during traversal. Inactive regions
will never even be seen by any ray, as they are not even in
the BVH being traversed.


The downside of this approach it that it requires updating the BVH
each time the transfer function or iso-value changes. However, even on our largest
data set (the \exajet), a full rebuild takes roughly 300ms on a Titan
RTX or RTX~8000 GPU. This time could be further improved by
refitting the BVH rather than rebuilding it.




\subsection{Adaptive Sampling}
\label{sec:adaptive}

Adaptive sampling is key to sampling the finest regions of
an AMR data set
at the same (relative) rate as the coarsest ones. For
example, on the \gear the coarsest cells are 4096 times the size of
the finest. Any technique that does not adapt the sampling rate by
the same factor will either grossly over-sample coarse regions or
under-sample fine ones.

To support adaptive sampling, we leverage each ABR's metadata to
adapt the sampling rate when traversing it to match the
frequency of the data it contains.
In contrast to Morrical et al.~\cite{morrical:19:adaptive},
we do not have to guess at the data frequency within a region, but can
simply take the smallest cell size within the region. We then set a
base sampling rate of two samples per smallest cell size in the
region. The sampling rate can be scaled
by a user-provided parameter to optionally subsample
the data to trade quality for rendering performance.
The examples shown in this paper use the high quality
base sampling rate.

%
For each region we compute the first
sample distance $t_0= dt (\rho + \lceil \frac{t0+\rho dt}{dt}
\rceil)$, where $\rho$ is a per-ray random offset for 
\emph{interleaved sampling}~\cite{keller::IS}, and $dt$ is
the scaled base sampling step. We then step the ray
through the region, sampling at each $t_{i+1}=t_i+dt$.
Our approach differs from the multi-block adaptive sampling
proposed by Ljung~\cite{ljung2006adaptive} in that we do
not define a global set of finest level samples that we skip
according to the coarseness of the region. Instead,
each region defines its own sampling intervals, independent
of the others.

\subsection{Opacity Correction}

\begin{figure}[t]
  \vspace{-0.5em}
  \centering
  \includegraphics[width=.75\columnwidth]{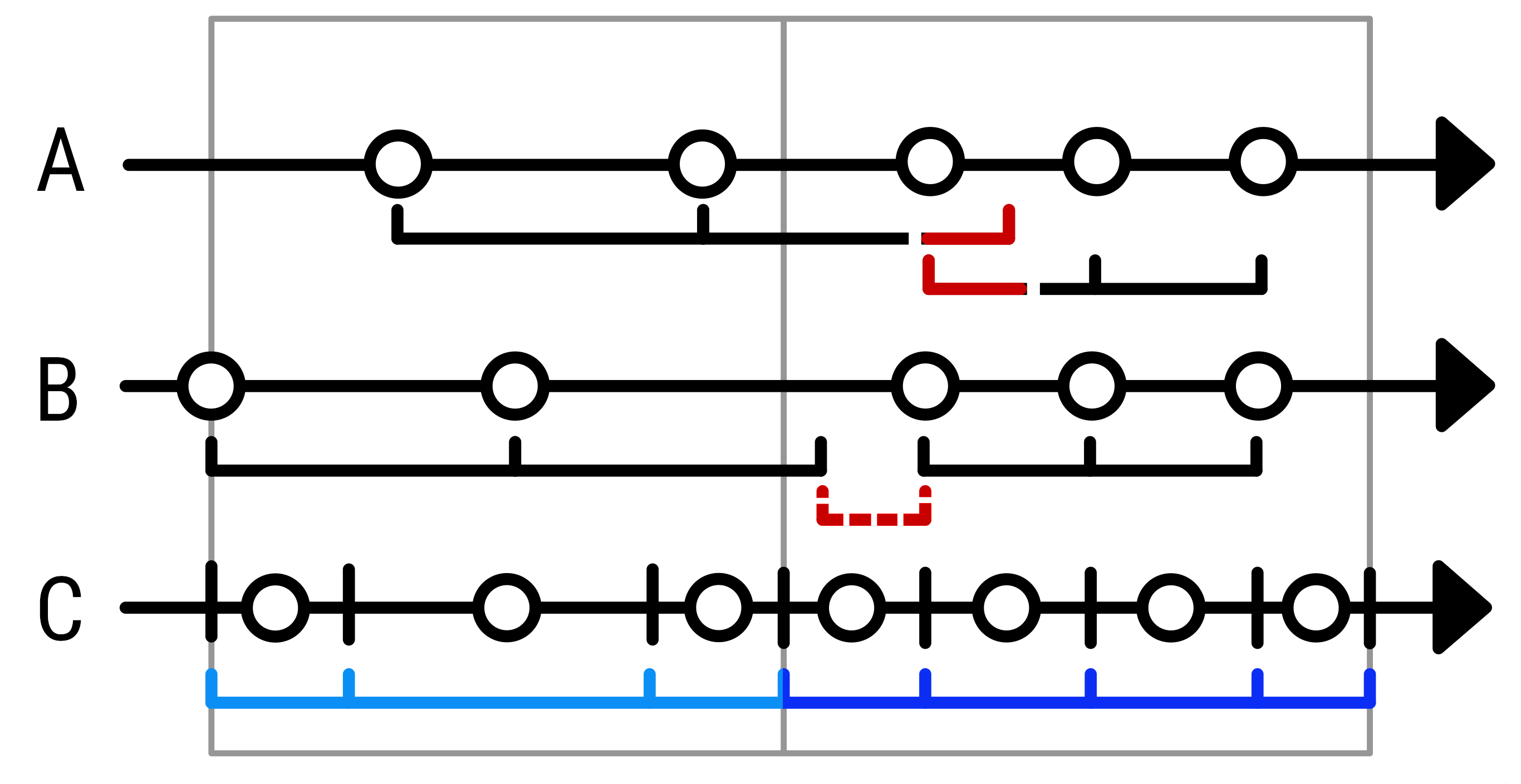}
  \vspace{-1em}
  \caption{\label{fig:opacity-correction}%
    If successive
    regions are sampled at different rates, the distance between the
    last sample in one and the first sample in the next can be
    smaller (A) or larger (B) than either region's sampling rate, leading
    to artifacts at region boundaries.
    We treat the sample points as delimiters which split
    the region's overlap interval into sub-intervals (C). Each sub-interval is
    sampled at its mid-point and weighed by its
    length, ensuring samples are spaced as expected.
    \vspace{-2em}
  }
\end{figure}

To account for the now variable step size between samples we use
opacity correction~\cite{engel_real-time_2006}, according to the term $\tilde{\alpha}=1-(1-\alpha)^{s/s_1}$,
where $s$ is the current step size, $s_1$ is the base step size, and $\alpha$ is the opacity
obtained from the transfer function.

However, even with this correction we encountered
rendering artifacts at region boundaries. We root-caused
these to the
fact that the first and last samples from successive regions can
lie closer together or further apart than the sample step
size of the adjoining regions would suggest (\Cref{fig:opacity-correction}).
To correct for this, we do not
actually sample at the sample positions $t_0,t_1,...t_n$, but
instead view them as delimiters for the intervals $T_0=[t_{in},t_0]$,
$T_1=[t_0,t_1]$, \dots, $T_{m}=[t_n,t_{out}]$. We then sample
each of these intervals in the middle
$\left(\frac{T_i[0]+T_i[1]}{2}\right)$, and
weigh it in the opacity correction term with the interval length,
$dt_i = |T_i|$.

\begin{figure}
    \vspace{-0.5em}
    \centering
    \begin{subfigure}{0.48\columnwidth}
        \centering
        \includegraphics[width=\textwidth]{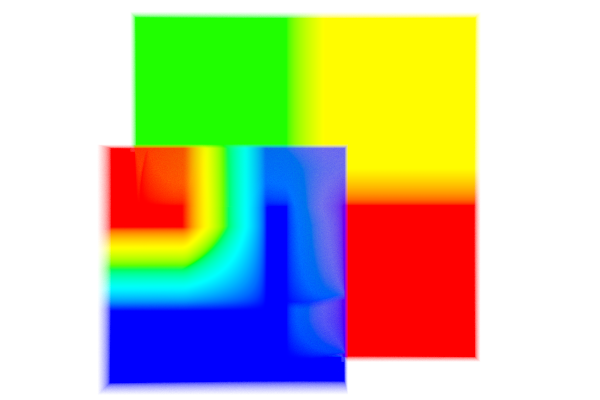}
    \vspace{-1.5em}
        \caption{w/o sample correction, $dt=0.3$}
    \end{subfigure}
    \begin{subfigure}{0.48\columnwidth}
        \centering
        \includegraphics[width=\textwidth]{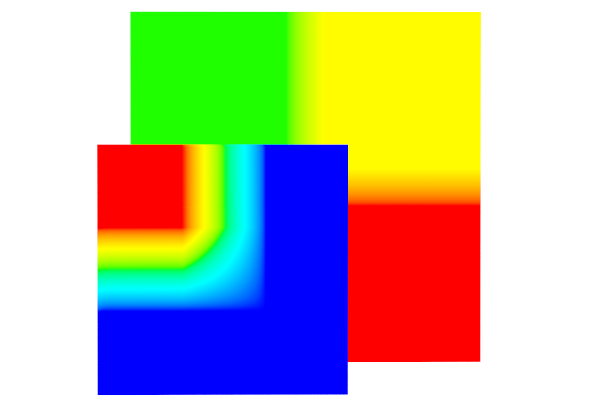}
    \vspace{-1.5em}
        \caption{w/ sample correction, $dt=0.3$}
    \end{subfigure}
    \begin{subfigure}{0.48\columnwidth}
        \centering
        \includegraphics[width=\textwidth]{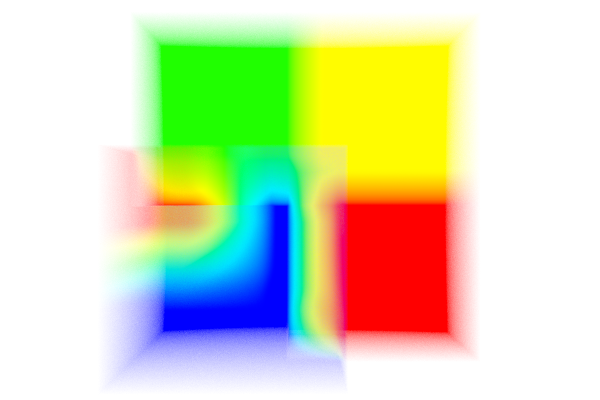}
    \vspace{-1.5em}
        \caption{w/o sample correction, $dt=2.0$}
    \end{subfigure}
    \begin{subfigure}{0.48\columnwidth}
        \centering
        \includegraphics[width=\textwidth]{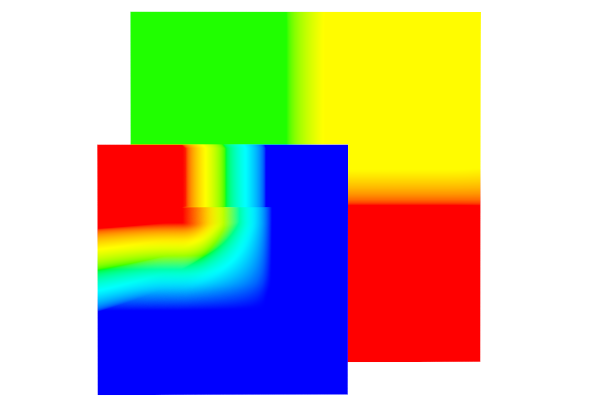}
    \vspace{-1.5em}
        \caption{w/ sample correction, $dt=2.0$}
    \end{subfigure}
    \vspace{-0.5em}
    \caption{\label{fig:correction}%
    We correct sample positions so that each region is always sampled
    at least once. Without correction, at low sampling
    rates (\emph{base} sample step $dt$),
    regions become translucent, while our sample correction
    manifests as less objectionable banding artifacts.
    Note that in the above images, the transfer function is fully opaque.
    \vspace{-2em}}
\end{figure}

Compared to the adaptive sampling approach described above, this correction takes one additional
sample from each region, and at least one sample for each region hit,
even if the overlap interval is small. With our approach the
opacity will not decrease due to undersampling, which can cause
objectionable artifacts. Instead, at lower sampling rates biasing
the sample positions results in some slight banding artifacts.
Although both
artifacts disappear in the limit with higher sampling rates, the
artifacts---and in fact incorrect opacity---from undersampling regions
are still very apparent at sampling rates where banding is
only very faint or not visible at all (see~\Cref{fig:correction}).

\ifx\empty
\subsection{Rendering the Active Brick Regions}

Once the active brick regions are built, using them
during
rendering works almost exactly as before, except that instead of
building the BVH over the bricks we build it over the active brick
regions. Otherwise the rendering code is exactly as before,
using ray tracing to step through successive regions and ray marching
within each region to integrate it. To compute a sample during the
ray marching process we iterate through the bricks active for the
current region (often just one) and perform \emph{basis} method reconstruction,
without any top-down traversal for cell queries whatsoever.


To support space skipping of active brick regions we store the
min/max scalar value of all cells whose support overlaps the active
brick region, then proceed as described in \Cref{sec:space-skipping}.
Similarly, for adaptive sampling each region stores the minimum cell
width of the bricks overlapping the region, but otherwise proceeds
as before.
\fi


\section{Implementation Details}
\label{sec:implementation}

We implement our method using OptiX 7~\cite{optix}, though we note
the same concepts are applicable to other GPU or CPU ray tracers.
All rendering operations are implemented in a \emph{ray generation program},
which performs both the iteration through the
bricks as well as the integration within each region. All data is
uploaded to CUDA memory buffers that this ray generation program
operates on. Multi-GPU rendering is supported by simply replicating
the data buffers on all GPUs, and assigning different GPUs to render different
regions of the image.


\subsection{Gradient Vectors}

Local shading with a bidirectional reflectance distribution function
(BRDF) requires gradient vectors to be computed as stand-ins for the
nonexistent surface normals. A standard approach for computing
these gradients is via central differencing at the sample point
(\Cref{sec:central-differencing}) at the cost of additional samples.
In our implementation, we employ an analytic gradient approach suitable for the \emph{basis}
reconstruction method that does not require additional samples (\Cref{sec:analytic-gradients}).

\subsubsection{Central Differencing}
\label{sec:central-differencing}

We compare two ways of computing central difference gradients. Both
require us to compute six additional basis reconstruction samples at positions
offset from the current sample at a distance proportional to the current region's
sample rate. When sampling at the boundary of a region these offsets may require
us to compute samples in other neighboring regions.
We perform the cell location similar to Wald et
al.~\cite{wald:19:tettyRTX}, by tracing an infinitesimal ray
originating at each offset position through the region BVH to find
the containing region.
Although the BVH traversal is hardware accelerated, this requires
six additional rays to be traced per sample, incurring significant
cost. 
To improve performance at the cost of quality, we also implemented a
variant that \emph{clamps} the offset positions to the current
region before evaluating them, removing the need for additional
ray traversals. 
We refer to this mode as
\emph{clamped central differences}. We observe that even with the
accurate method, gradients are not necessarily continuous, as the
offset size can change at level boundaries.

\subsubsection{Analytic Gradients}
\label{sec:analytic-gradients}
\begin{figure}[t]
    \begin{center}
    \includegraphics[width=0.8\columnwidth]{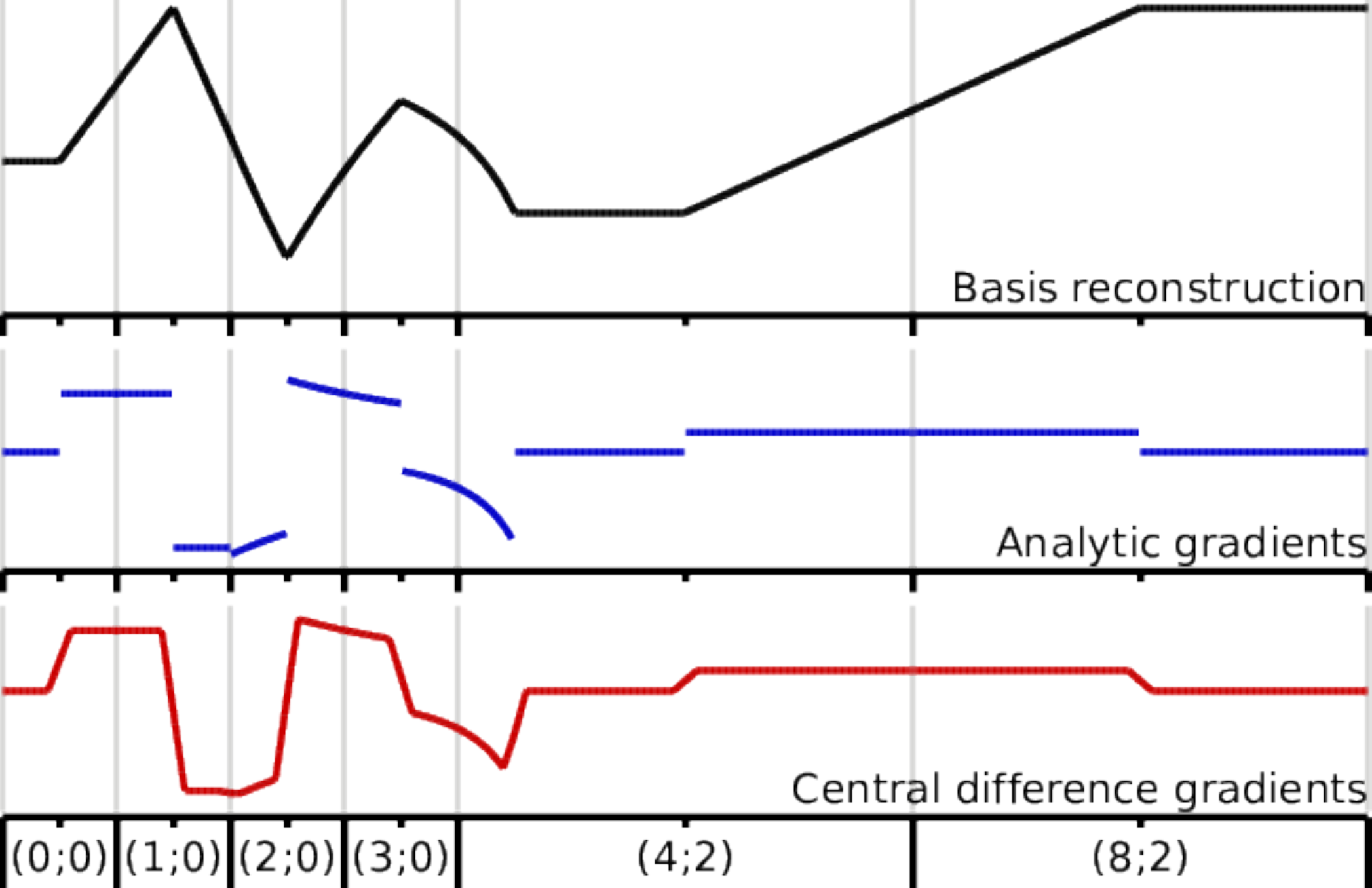}
    \end{center}
    \vspace*{-1.5em}
  \caption{\label{fig:gradients1D}%
  A 1D illustration of analytic vs.\ central difference gradients.
    Top: basis reconstruction
  with four level-0 and two level-2 cells. Middle: derivative of the
  reconstructed signal computed analytically from the same data values
  used for the reconstruction. Bottom: central differences
  require the signal to be
  reconstructed at additional positions to the left and
  right of the sample. Basis reconstruction
  is continuous, but not continuously differentiable,
  leading to shading artifacts at
  cell and level boundaries. Central difference derivatives are
  continuous but expensive to compute.
    \vspace*{-1.75em}}
\end{figure}
\begin{figure}[ht]
    \centering
    \vspace{-0.25em}
    \includegraphics[width=0.85\columnwidth]{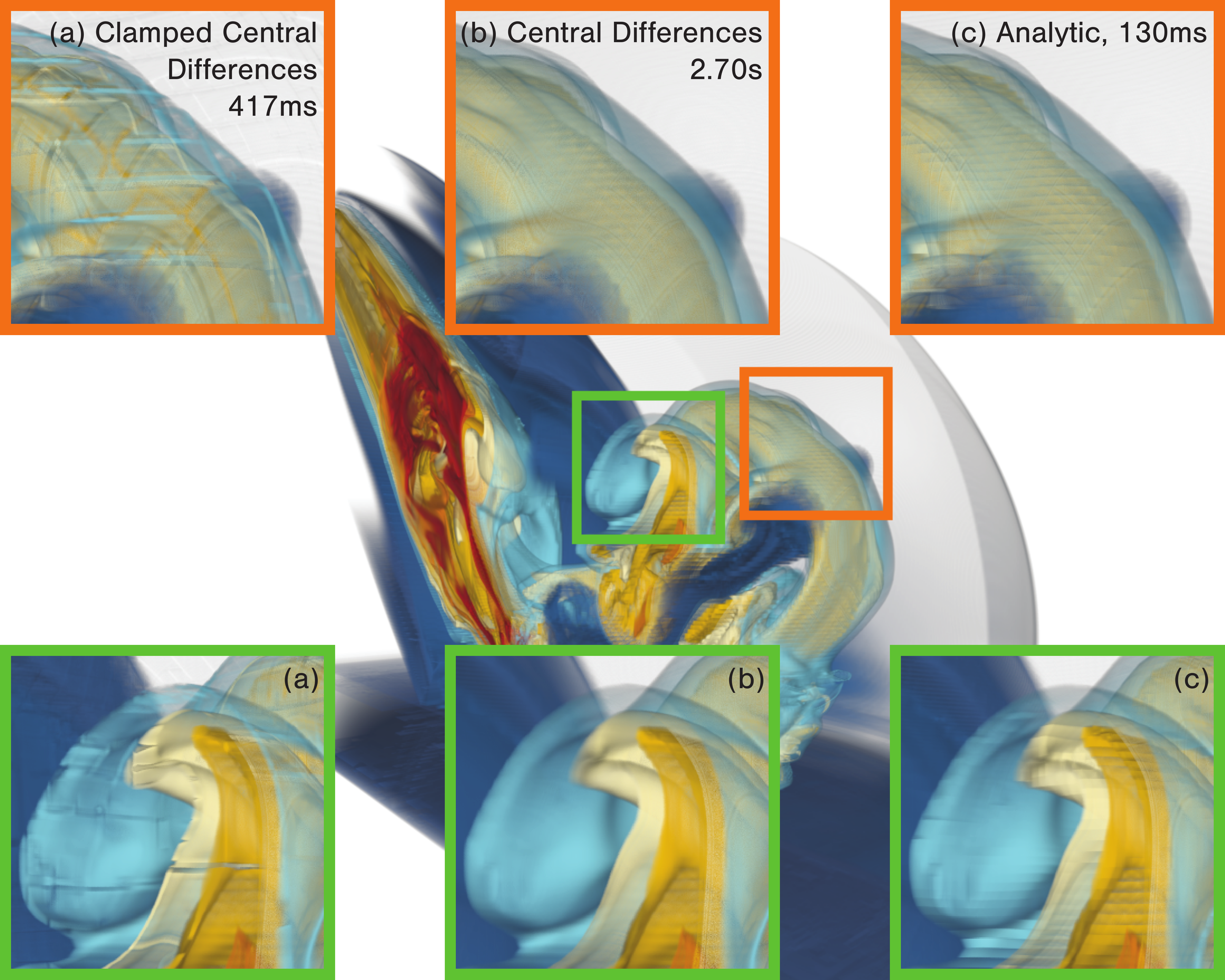}
    \vspace{-0.5em}
    \caption{\label{fig:gradients}%
    Different methods to compute gradients for shading, and rendering
    times obtained for the \impact data set.
    (a) Clamped Central Differences: sample positions are clamped
    to remain within the region for performance, resulting in artifacts
    at region and level boundaries.
    (b) Central Differences: samples in neighboring regions required
    to compute the central difference are looked up through additional
    ray traversals. The resulting gradients are high-quality but
    prohibitively slow to compute.
    (c) Analytic: when computing the basis function
    contribution we determine the partial derivatives analytically
    using \autoref{eq:deriv}. Analytic gradients are nearly as accurate
    as (b), but do not require additional data loads or ray traversal,
    and thus incur little performance impact.
    \vspace{-1.25em}
    }
\end{figure}

For our third option to compute gradients, we observe that for the
basis reconstruction method
it is actually possible to compute the gradient analytically.
This means that the gradient can be computed using just the existing
data values loaded for the original sample evaluation, without
additional memory accesses or ray traversals.
The gradient can be computed as the first order partial derivatives
of \Cref{eq:basis,eq:hat} (shown only for $x$ below as an example):
{\relsize{-1}{
\begin{equation}
  \label{eq:deriv}
  \frac{
    \partial B(p)
  }{\partial x} =
  \frac{\sum_{C}\Test{H_{C}}(p)\sum_{C}\frac{\partial
      \Test{H_{C}}(p)}{\partial x}C_{v} -
    \sum_{C}\Test{H_{C}}(p)C_{v}\sum_{C}\frac{\partial
    \Test{H_{C}}(p)}{\partial x}}
       {(\sum_{C}\Test{H_{C}}(p))^2},
\end{equation}}}
with
{\relsize{-1}{
\begin{equation*}
  \frac{\partial \Test{H_{C}}(p)}{\partial x} = \Test{h}\bigg(
  \frac{|C_{p_y}-p_y|}{C_w}\bigg) \Test{h}\bigg(
  \frac{|C_{p_z}-p_z|}{C_w}\bigg) \chi(x)\frac{1}{C_w}
\end{equation*}}}
and
{\relsize{-1}{\begin{equation*}
  \chi(t) = \begin{cases}
    -1, & \text{if } C_{p_{t}} - p_{t} \geq 0\\
    1,  & \text{otherwise}
  \end{cases}
\end{equation*}}}

Central differences will just connect neighboring point samples in an
epsilon region and are thus guaranteed to be continuous. In
comparison, the analytic gradient is arguably more ``correct'' but
not always continuous (see \Cref{fig:gradients1D}), which can
lead to slightly worse image quality compared to
central differences. However, analytic gradients provide far
superior quality than clamped central differences
and are much faster to compute than both central difference
variants (\Cref{fig:gradients}). Thus, they are used by default in our renderer.



\subsection{Rendering Modes}


\label{sec:bvh-iso}

\begin{figure}[ht!]
  \begin{center}
    \resizebox{0.9\columnwidth}{!}{
      \includegraphics[height=6cm]{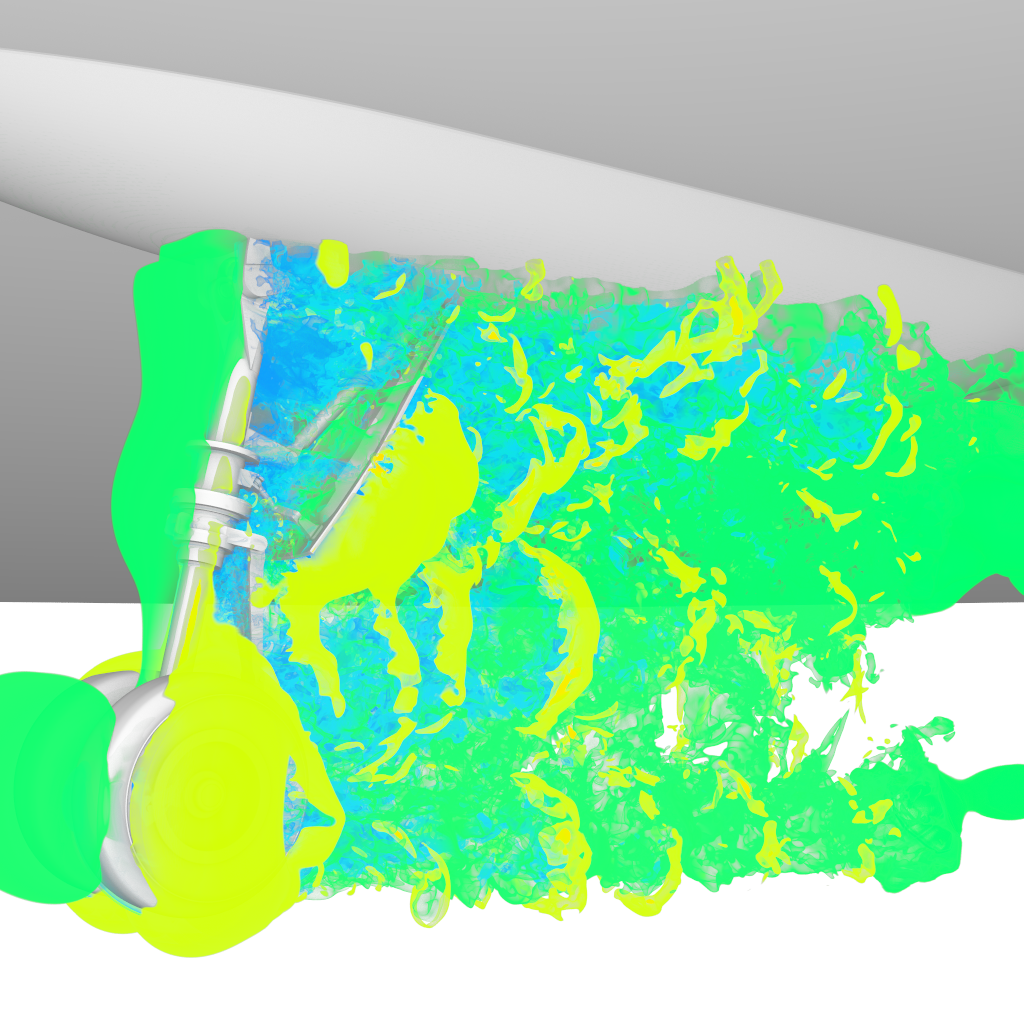}
      \includegraphics[height=6cm]{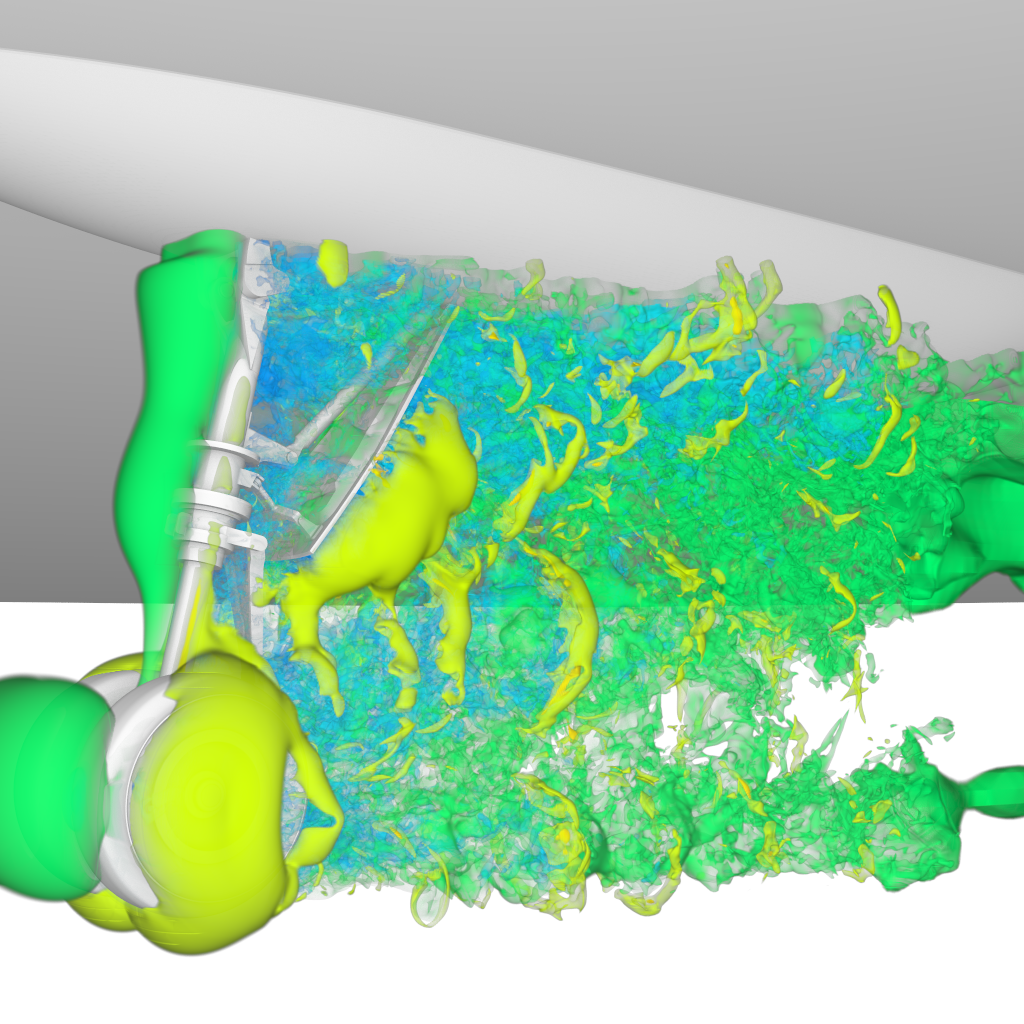}
      \includegraphics[height=6cm]{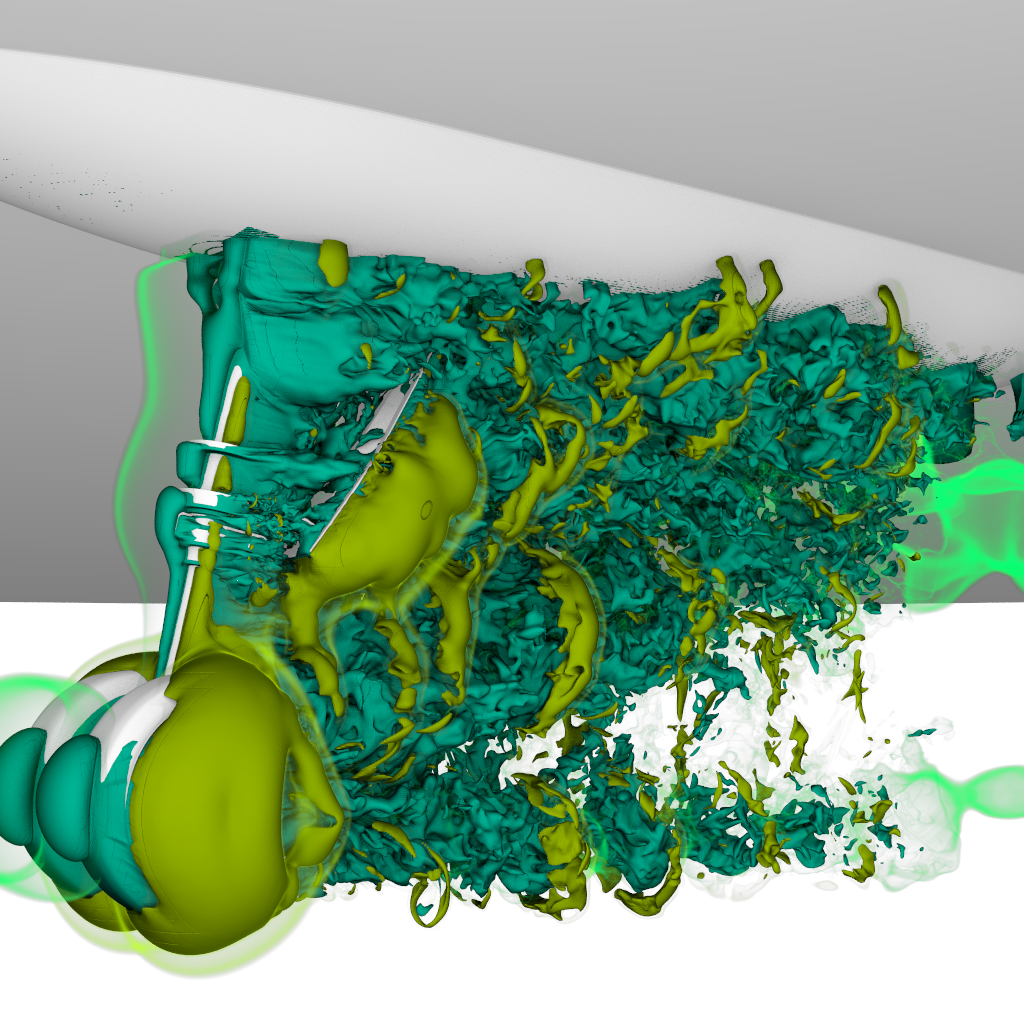}
    }
    \resizebox{0.9\columnwidth}{!}{
      \includegraphics[height=6cm]{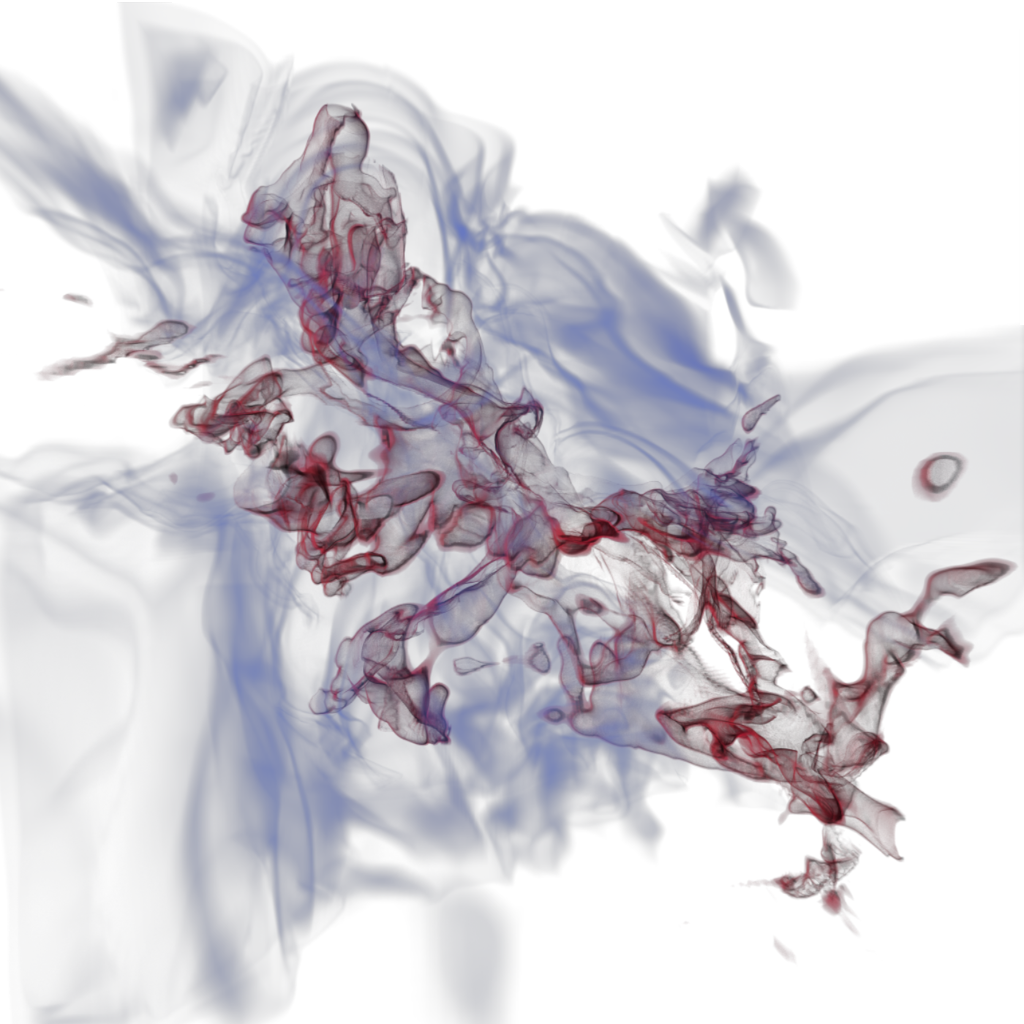}
      \includegraphics[height=6cm]{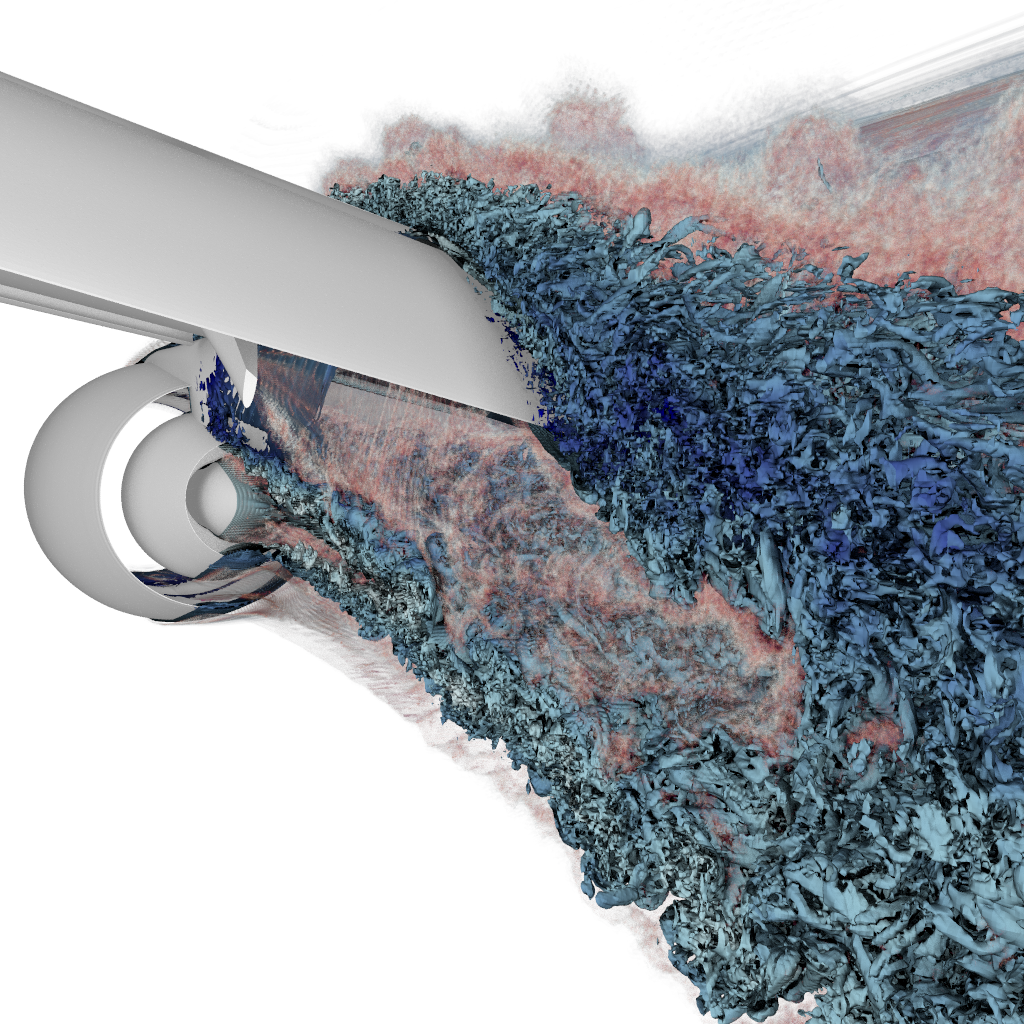}
    }
  \end{center}
    \vspace{-1.5em}
  \caption{\label{fig:effects}%
    Examples of the rendering features in our framework.
    Top: the \gear with direct volume rendering only (left),
    gradient shading using analytic gradients (center), and
    with implicit iso-surface, volume rendering, and ambient
    occlusion (right).
    Bottom: Multi-field volume rendering of the \cloud (left),
    and the \exajet with a triangle surface, color-mapped iso-surface, and volume,
    clipped to highlight interior features, with ambient occlusion.}
    \vspace*{-1.5em}
\end{figure}

\begin{figure*}
  \setlength\tabcolsep{.5ex}
  \centering
  \begin{tabular}{cccc}
      \includegraphics[width=.24\textwidth]{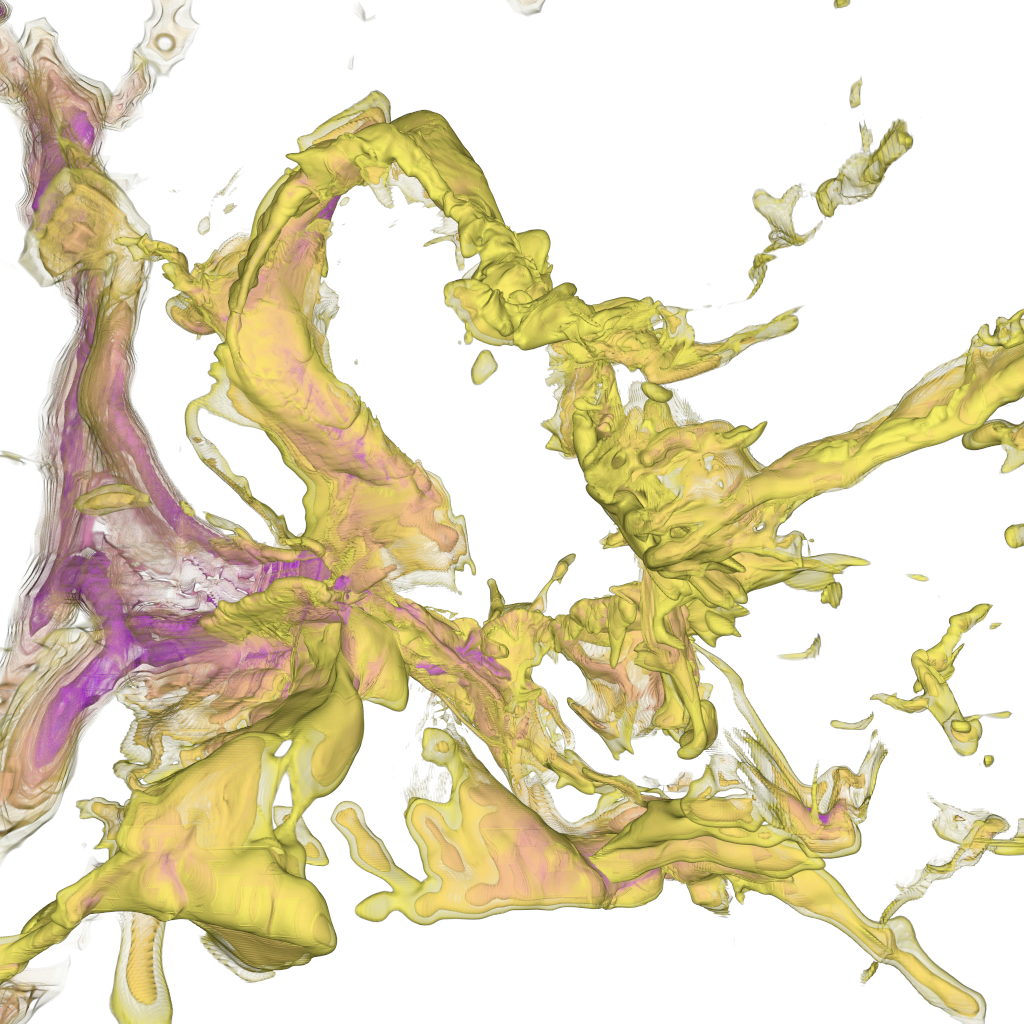}&
      \includegraphics[width=.24\textwidth]{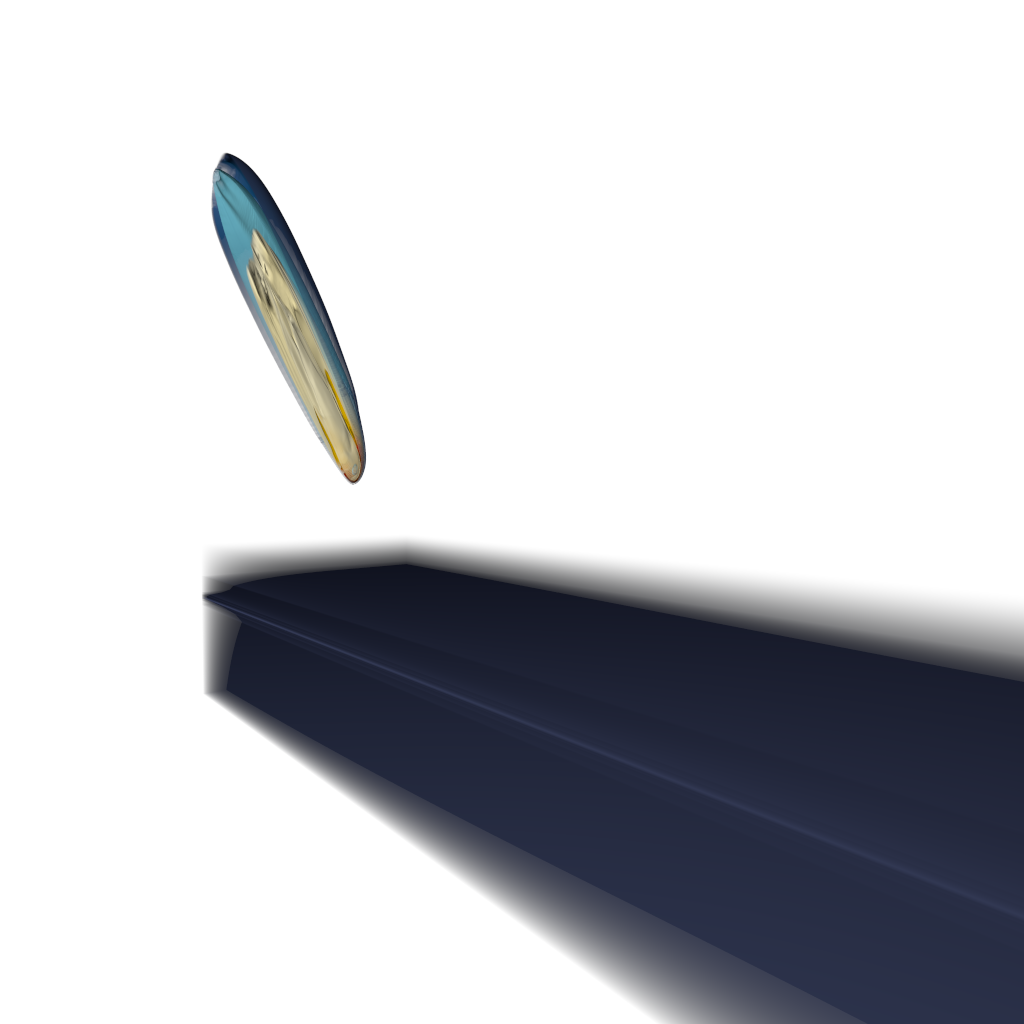}&
      \includegraphics[width=.24\textwidth]{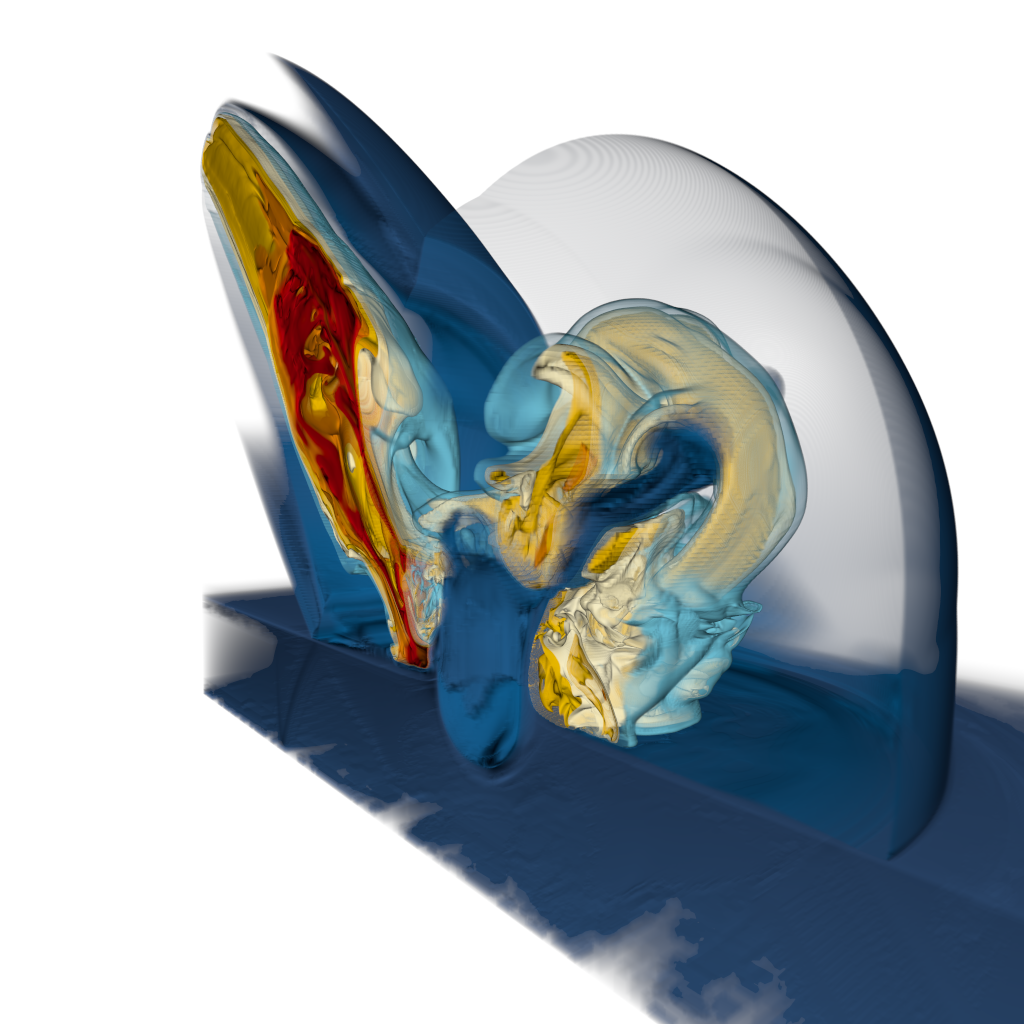}&
      \includegraphics[width=.24\textwidth]{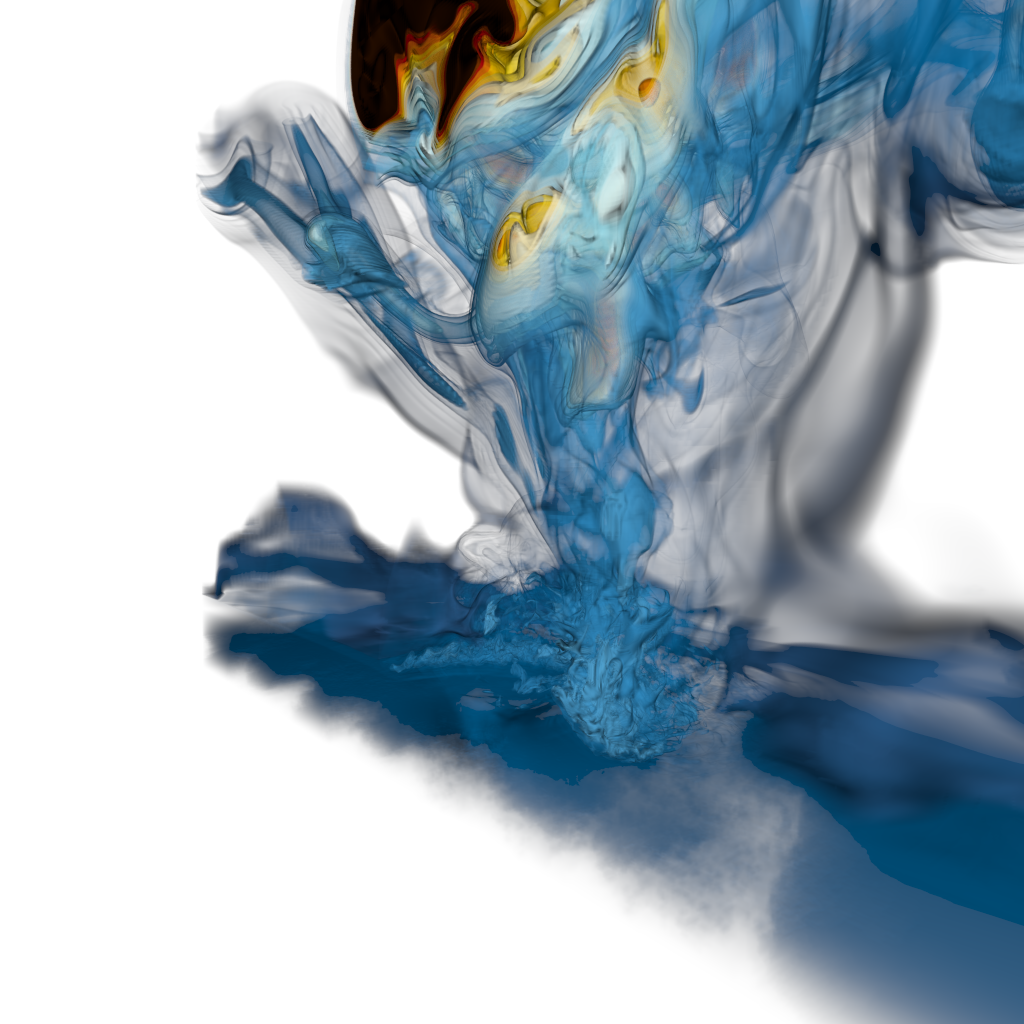}
      \\
      \textsf{\small \cloud}
      &
      \textsf{\small \impact (t=5700)}
      &
      \textsf{\small \impact (t=20060)}
      &
      \textsf{\small \impact (t=46112)}
      \\
      \textsf{\small 102M cells, 62.5~ms} 
      &      
      \textsf{\small 26.8M cells, 51.5~ms} 
      &      
      \textsf{\small 158M cells, 69.6~ms} 
      &      
      \textsf{\small 283M cells, 153.5~ms} 
      \\
      \includegraphics[width=.24\textwidth]{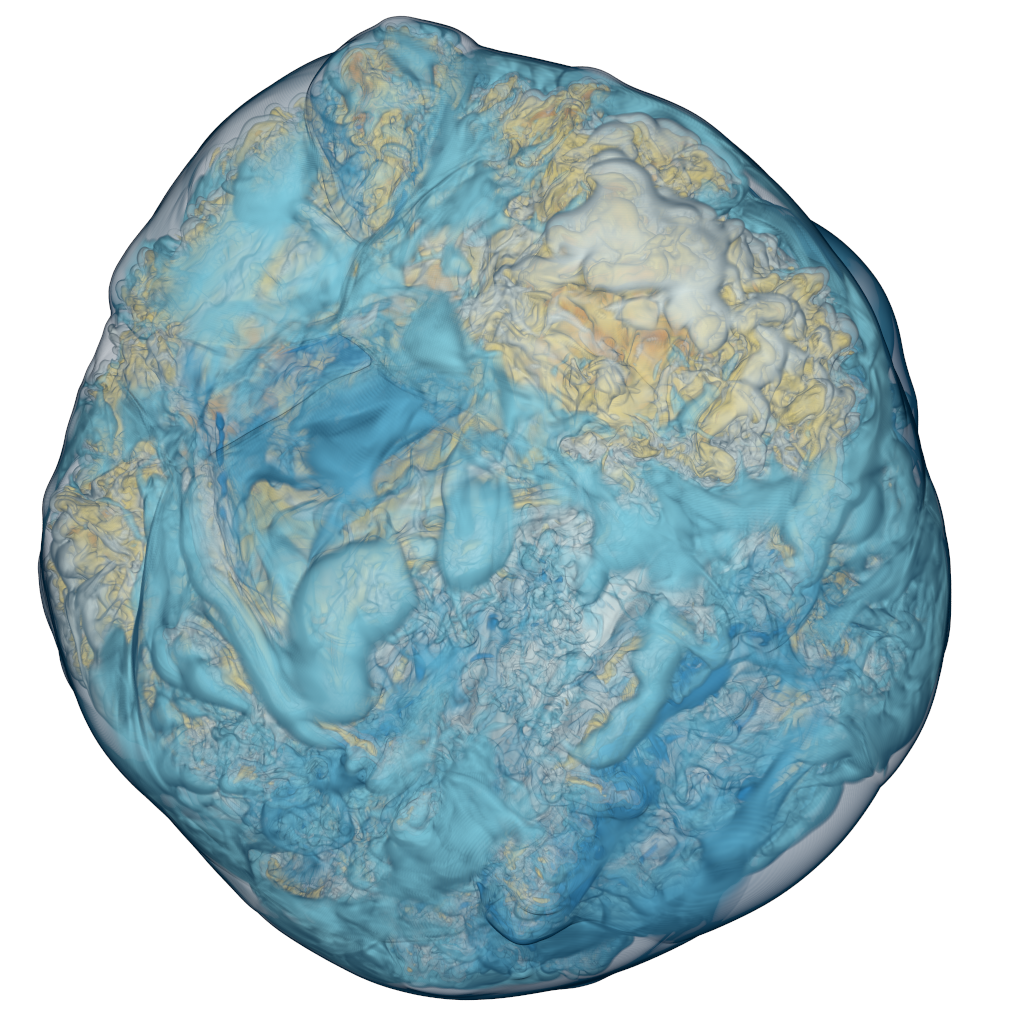}&
      \includegraphics[width=.24\textwidth]{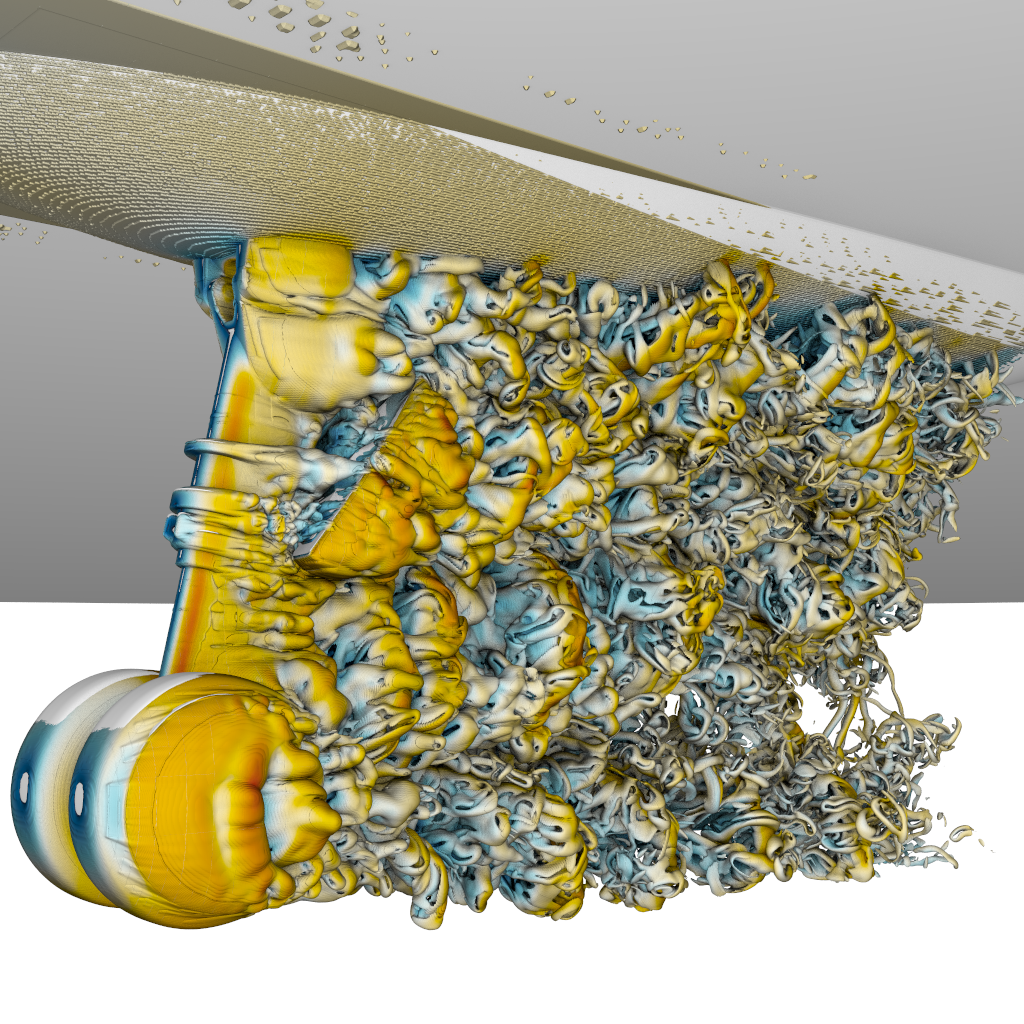}&
      \includegraphics[width=.24\textwidth]{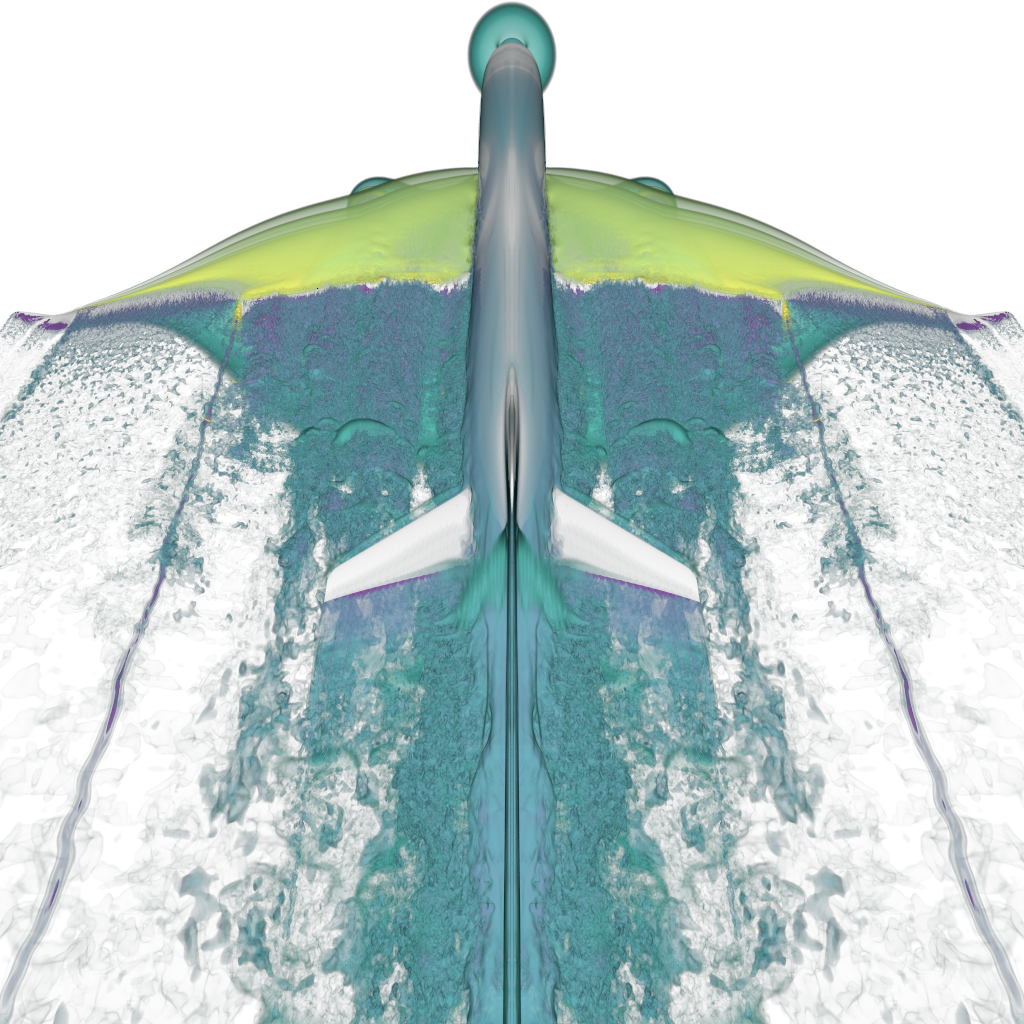}&
      \includegraphics[width=.24\textwidth]{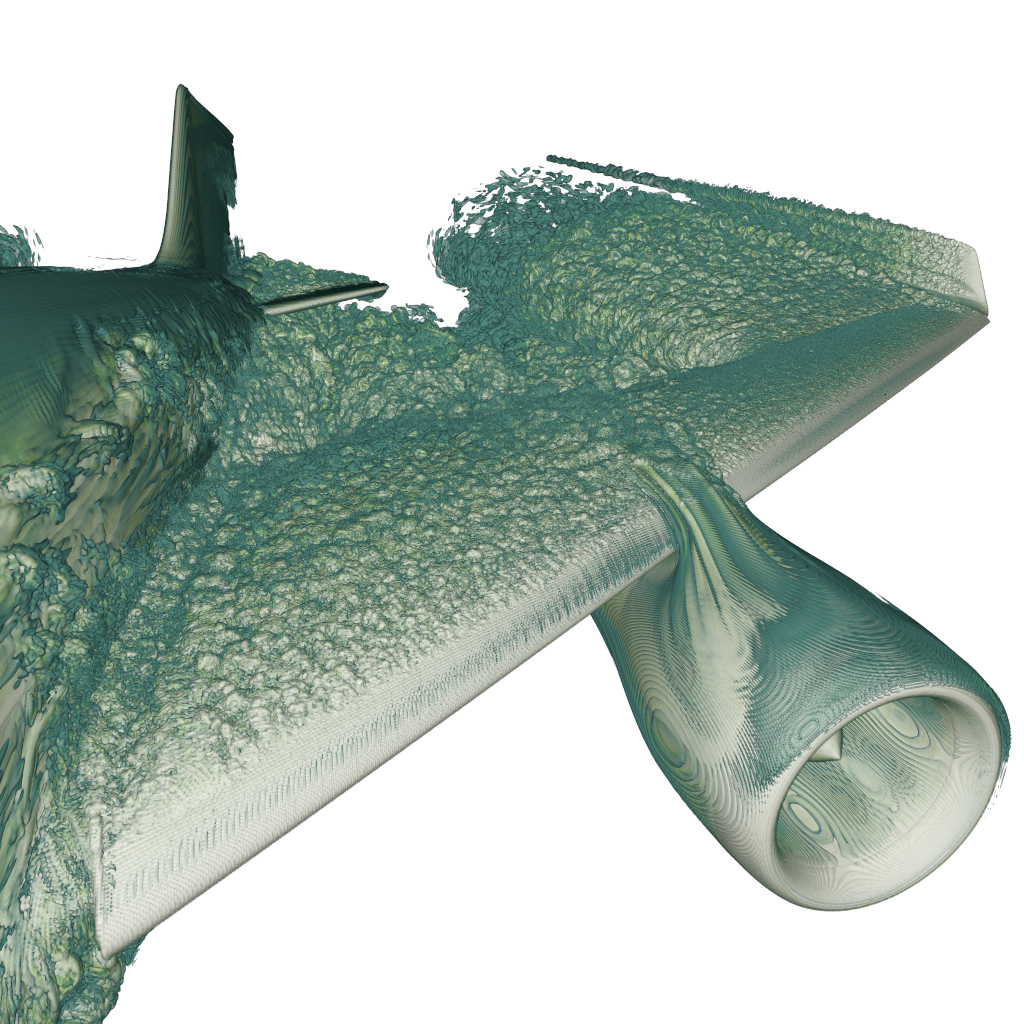}
      \\
      \textsf{\small \stellar}
      &
      \textsf{\small \gear (iso-surface)}
      &
      \textsf{\small \exajet (rear, velocity)}
      &
      \textsf{\small \exajet (wing, vorticity)}
      \\
      \textsf{\small 411M cells, 61.1~ms}
      &      
      \textsf{\small 262M cells, 1.59M tris., 71.8~ms} 
      &      
      \textsf{\relsize{-1}{1.31B cells, 126M tris., 80.7~ms}} 
      &      
      \textsf{\relsize{-1}{1.31B cells, 126M tris., 60.1~ms}} 
  \end{tabular}
  \vspace{-1em}
  \caption{\label{fig:models}%
    The data sets and benchmark visualizations used in evaluating our method.
    Render times are reported on a
    workstation with two RTX~8000 GPUs at 1024$\times$1024 pixels
    using the highest sampling quality settings and two AO rays
    per-pixel. For the LANL Impact we applied volume clipping
    through the middle of the data to see the plume's interior. The Exajet is mirrored
    down the middle of the plane, the original data contains 656M cells and 63.2M triangles.
    \vspace{-2em}
  }
%
      %
\end{figure*}

We support a variety of different rendering modes to be able to evaluate
our data structure under realistic conditions, including implicit
iso-surfaces and direct volume rendering using ray marching. We
maintain
separate BVHs for volume data and iso-surfaces, as iso-surfaces
in general have much more potential for empty space skipping due to
their sparsity. As with the volume BVH, the iso-surface BVH must be
rebuilt when the iso-value changes. We also support surface
geometry represented as triangle meshes which are rendered using
the RTX hardware-accelerated ray-triangle intersection test.
We implement clipping planes by setting the rays'
\code{[ray.tmin, ray.tmax]} intervals accordingly (demonstrated on the
\exajet in \Cref{fig:effects} and the \impact in \Cref{fig:models}).

In the presence of iso-surfaces or meshes, we first trace each ray
against the (fully hardware-accelerated) mesh BVH, then transform it
into the volume space, and trace it against the iso-surface BVH to
check for a closer surface hit point. We then shorten the ray
to the nearest surface hit point, if any, and trace it through the
volume BVH to integrate the volume up to that point. For better depth
cues, we also support ambient occlusion rendering in addition to
local shading.

Bricks and regions contain only spatial information about the cells,
but can refer to more than one scalar field. We exploit this property
to support color-mapping an iso-surface computed on one field with
colors computed from another---for example, flow velocity mapped on to the
vorticity iso-surface in \Cref{fig:teaser}. We also implemented
a multi-field volume renderer where each sample point's color and
opacity is computed as the combination of different scalar fields'
independent transfer functions. Example renderings with the various
supported modes are shown in \Cref{fig:effects}.

\section{Results}
\label{sec:results}

To evaluate our method we performed a set of benchmarks on a range of
medium to large AMR data sets (\Cref{fig:models}). Our
benchmarks are performed on a workstation with an Intel Xeon CPU (8 cores, 2.2~GHz),
128~GB RAM, and two NVIDIA RTX~8000 GPUs, each of which has 4608 CUDA
cores, 72 RT Cores, and 48~GB of GDDR6 VRAM\footnote{We
also ran our benchmarks on Titan~RTX GPUs, with similar results.}.
We use Ubuntu 18.04, NVIDIA driver 440.44, OptiX 7.0, and CUDA 10.2.


Unless otherwise mentioned, all benchmarks are performed at the highest
quality settings; i.e., using the basis method interpolant, per-sample
gradient shading, surface geometry (where provided), ambient occlusion
with two rays per-pixel on surface and iso-surface geometry, and an
integration step size of two samples per cell.
Implicit iso-surfaces are mentioned explicitly where used.
The benchmarks are run using our interactive viewer, where the
visualization parameters can be modified
interactively by the user.

\subsection{Data Sets}
\label{sec:data-sets}

The structure and complexity of AMR data can vary widely among
different data sets and codes. To provide a representative set of
benchmarks we spent significant effort to
cover a range of formats, codes, and
model complexity (\Cref{fig:models}).

\textbf{TAC Molecular Cloud} and \textbf{Princeton Stellar
Cluster Wind} are astrophysics simulations from the
Theoretical Astrophysics Group in Cologne~\cite{seifried_2017} and
Princeton~\cite{soares_wind_2019},
respectively; computed with the Flash simulation
code~\cite{Fryxell_2000}. Flash comes with a designated HDF5-based
file format that stores the simulation grid and multiple simulation
output variables in a single file, from which we extracted the AMR
leaf cells.

\textbf{LANL Deep Water Impact} is a simulation
of an Asteroid Ocean Impact computed with
xRage~\cite{xrage} (see Patchett et al.~\cite{impact-dataset}).
Of particular interest for this data set is that the entire
time series is available, and that the AMR structure is refined over
time.
\Cref{fig:models}).

\textbf{Landing Gear} is the same data set used in previous
AMR rendering research by Wald et al.~\cite{wald:17:AMR} and Wang et
al.~\cite{wang:18:iso-amr}. The data set is a simulation of air flow
around an airplane's landing gear, simulated with NASA's LAVA
code~\cite{lava}. To import this data
set we first loaded it into OSPRay~\cite{ospray}, and modified
OSPRay's AMR module to iterate over its AMR $k$-d tree's leaf nodes and
write out the contained cells. Of particular interest for this
data set is the large ratio of the coarsest to finest cell sizes, at
$4096:1$. The Landing Gear also includes 1.59M triangles in surface
geometry.

\textbf{Exajet} is a simulation of air flow around a jet
plane~\cite{casalino2014lattice}
performed using PowerFLOW~\cite{powerflow}. The model contains 656M
cells across four levels; its finest level covers a logical grid
of
$4.8K\times 2.4K\times 2.1K$. Of particular interest for this data set
is that the interior of the airplane is not covered with cells,
resulting in curved finest-level cell boundaries along the
fuselage and wings. The Exajet also includes 63.15M triangles of surface
geometry. As the data is cut along the symmetry down the middle of the
fuselage, we create an additional instance mirrored along this axis to
produce a visualization of a complete jet. After instancing, the scene
contains 1.31B cells and 126M triangles.

\subsection{Memory Consumption}

We instrumented our code to
track the sizes of the individual scalar, brick, region, and triangle
data arrays, and measured final memory usage using the \code{nvidia-smi}
tool (see~\Cref{tab:memory}). We note that this does not capture some
 additional temporary memory that OptiX uses during BVH construction.
Memory consumption---and in particular, the memory used by OptiX---varies
widely across the different data sets; however, even the most complex
models easily fit into a Titan RTX's or RTX 8000's GPU memory, even with multiple fields.




\begin{table}
    \caption{\label{tab:memory}%
    GPU memory use for the benchmarks shown in \Cref{fig:models}.
    ``Total'' was measured using
    \code{nvidia-smi} and includes auxiliary data such as
    the frame buffer, accumulation buffer, and BVH memory.}
    \vspace{-1em}
  \centering
  \relsize{-1}{
    \begin{tabular}{@{}lrrrrr@{}}
    \toprule
    & \multicolumn{3}{c}{Volume Data}  & Surface Data & Total\\
        \cmidrule(lr){2-4}
    Model & Scalars & Bricks & Regions &  & \\
    \midrule
        Cloud & 307MB & 31.8MB & 1.00GB & n/a & 2.06GB \\
        Impact-5K & 102MB & 17.7MB & 676MB & n/a & 2.19GB \\
        Impact-20K & 604MB & 104MB & 4.82GB & n/a & 12.9GB \\
        Impact-46K & 1.06GB & 95.1MB & 4.15GB & n/a & 11.7GB \\
        Wind & 1.53GB & 767KB & 36.3MB & n/a & 2.16GB \\
        Gear & 1.96GB & 813KB & 42.2MB & 38.2MB & 2.70GB \\ 
        Exajet & 2.45GB & 95.0MB & 2.95GB & 1.52GB & 13.4GB \\ 
    \bottomrule
    \end{tabular}
  }
  \vspace{-2em}
\end{table}

\subsection{Performance}

\begin{figure}[t]
    \centering
    \vspace{-1em}
    \begin{subfigure}{0.48\columnwidth}
        \centering
        \includegraphics[width=\textwidth]{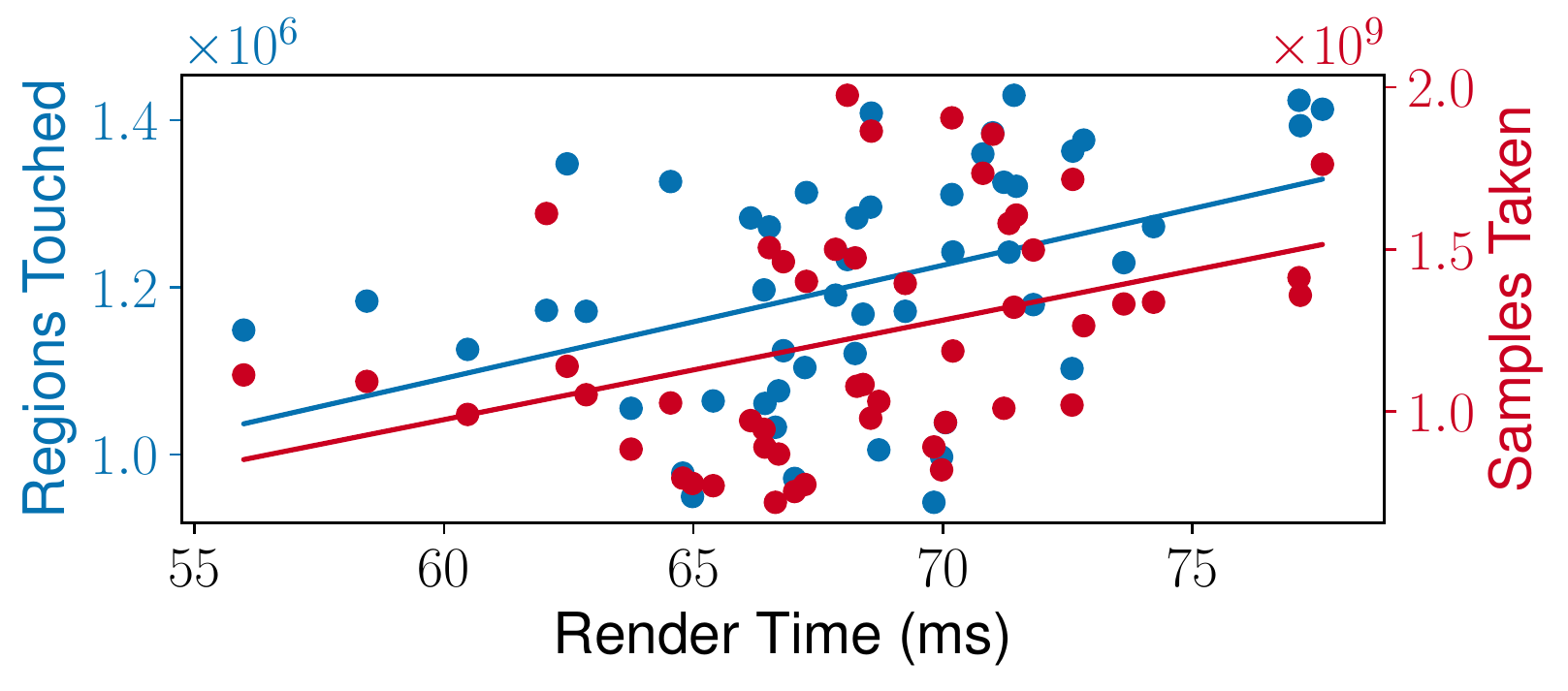}
        \vspace{-1.75em}
        \caption{\label{fig:molecular_cloud_perf_v_regions}%
        TAC Molecular Cloud}
    \end{subfigure}
    \begin{subfigure}{0.48\columnwidth}
        \centering
        \includegraphics[width=\textwidth]{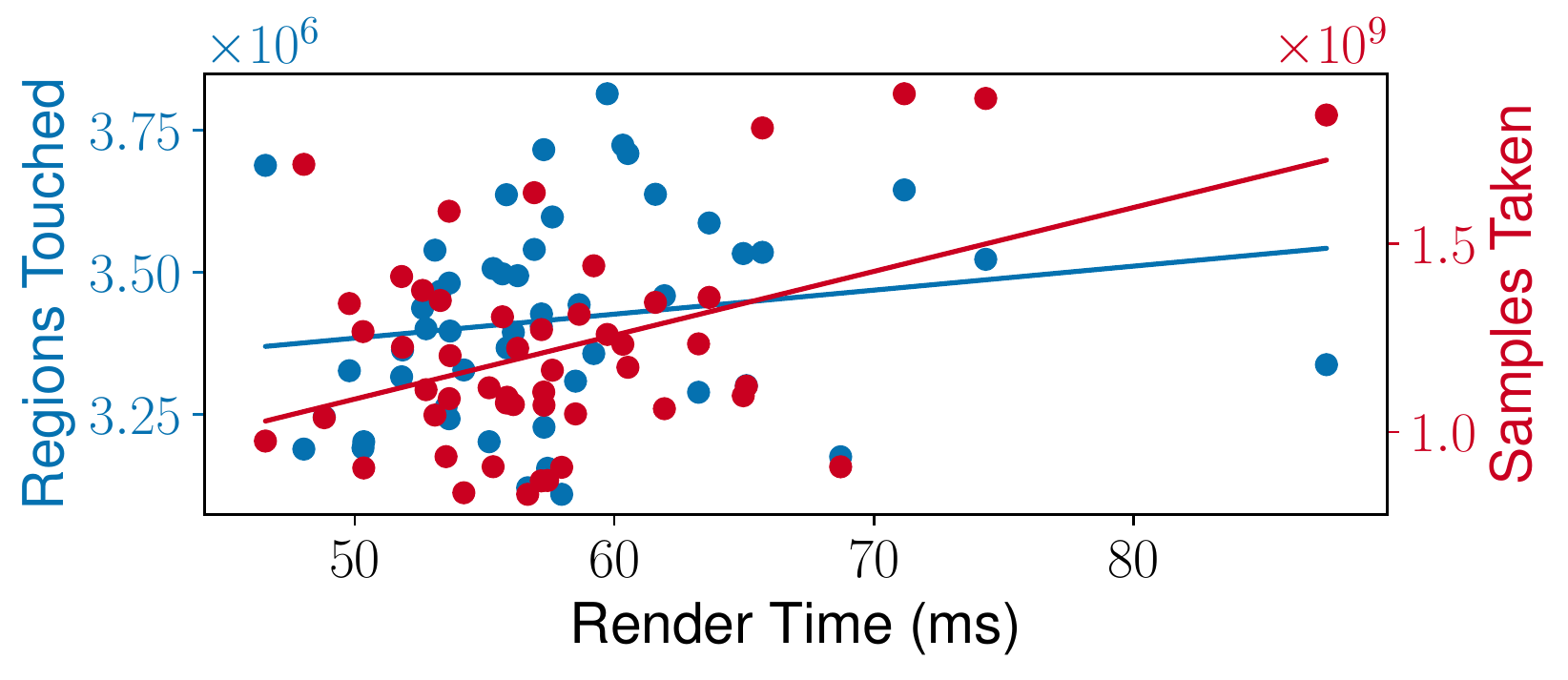}
        \vspace{-1.75em}
        \caption{\label{fig:meteor_bench_5k_perf_v_regions}%
        LANL Impact (t=5700)}
    \end{subfigure}
    \begin{subfigure}{0.48\columnwidth}
        \centering
        \includegraphics[width=\textwidth]{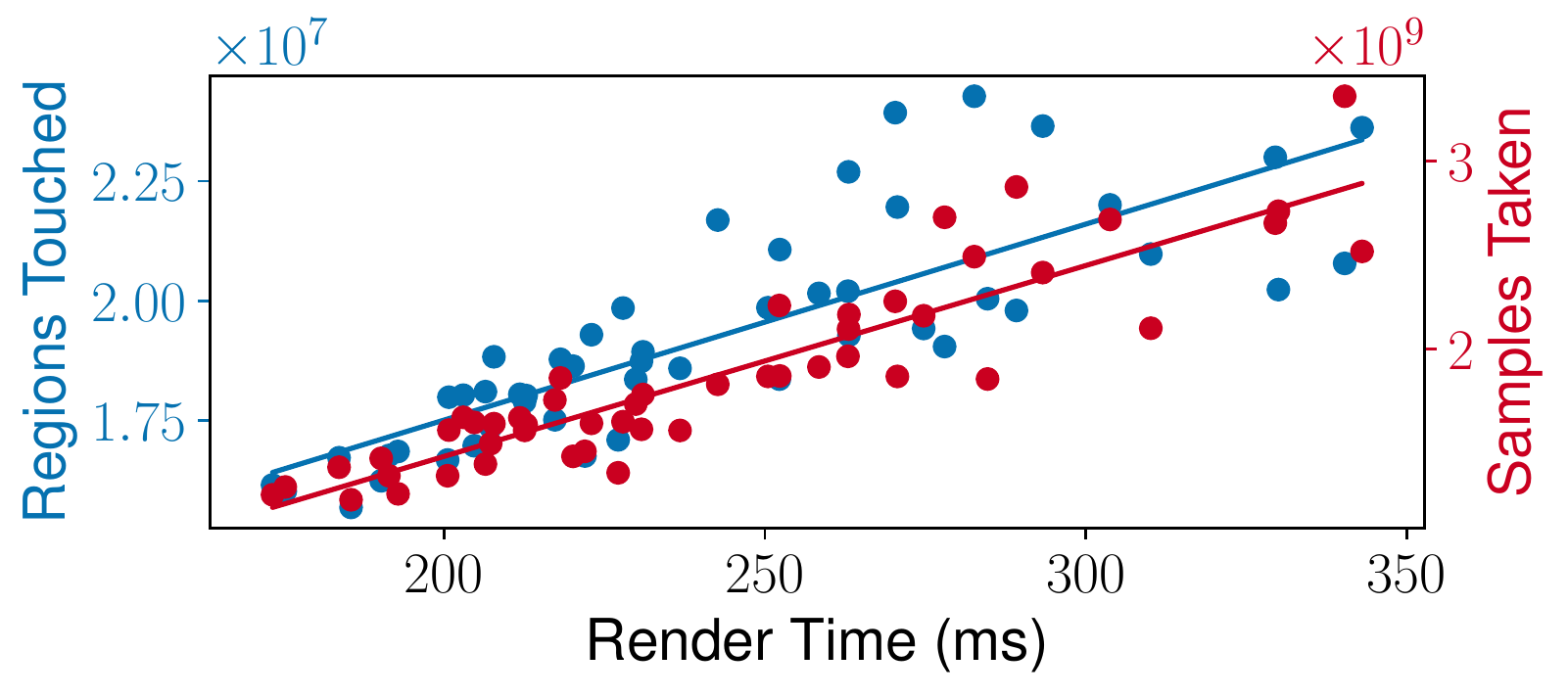}
        \vspace{-1.75em}
        \caption{\label{fig:meteor_bench_20k_perf_v_regions}%
        LANL Impact (t=20060)}
    \end{subfigure}
    \begin{subfigure}{0.48\columnwidth}
        \centering
        \includegraphics[width=\textwidth]{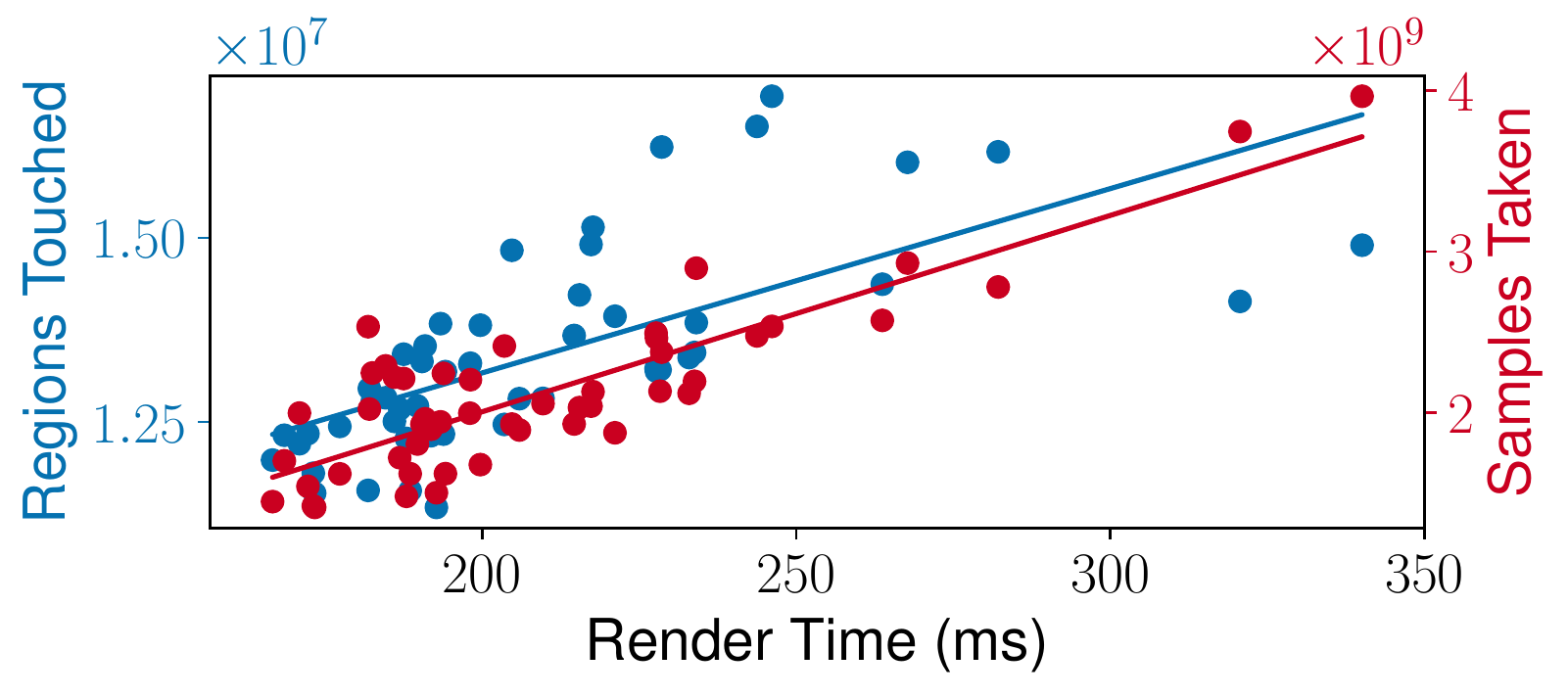}
        \vspace{-1.75em}
        \caption{\label{fig:meteor_bench_46k_perf_v_regions}
        LANL Impact (t=46112)}
    \end{subfigure}
    \begin{subfigure}{0.48\columnwidth}
        \centering
        \includegraphics[width=\textwidth]{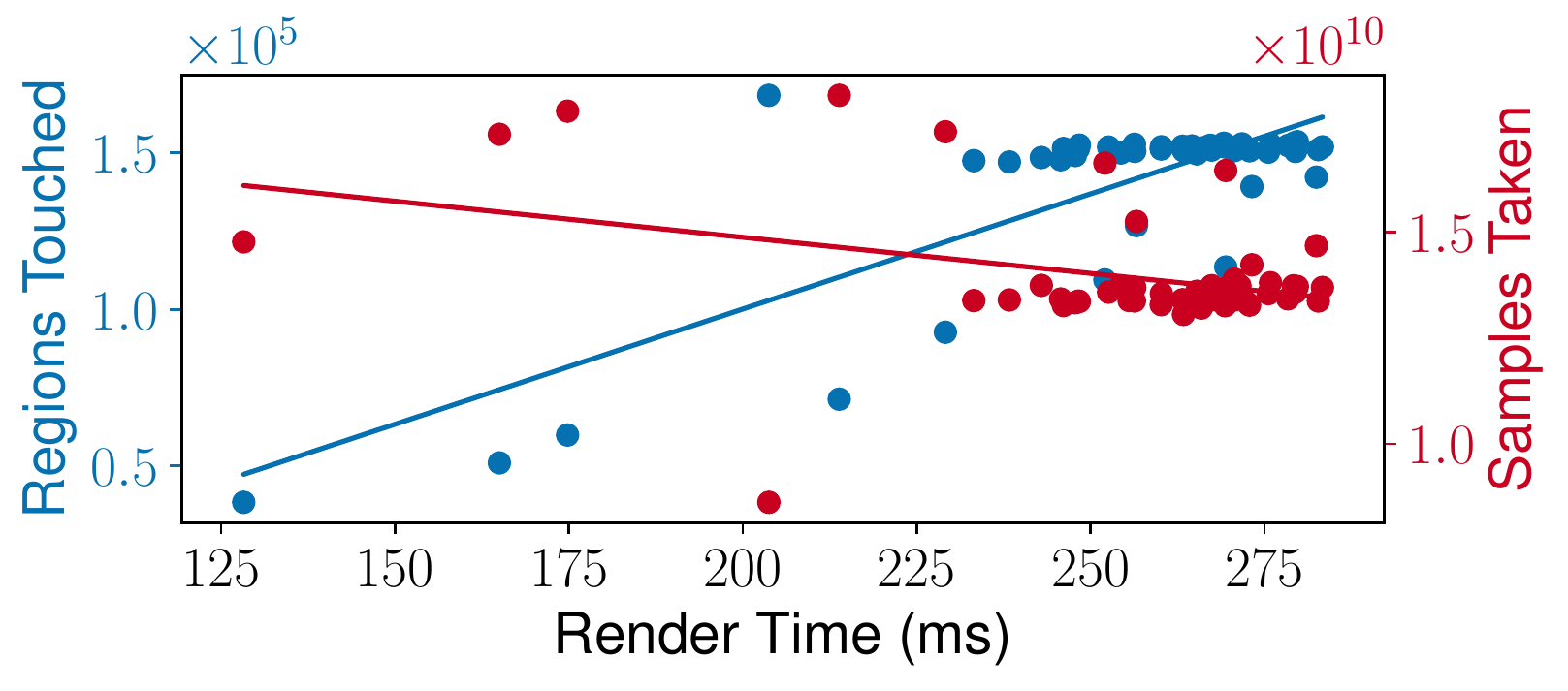}
        \vspace{-1.75em}
        \caption{\label{fig:princeton_perf_v_regions}%
        Princeton Stellar Cluster Wind}
    \end{subfigure}
    \begin{subfigure}{0.48\columnwidth}
        \centering
        \includegraphics[width=\textwidth]{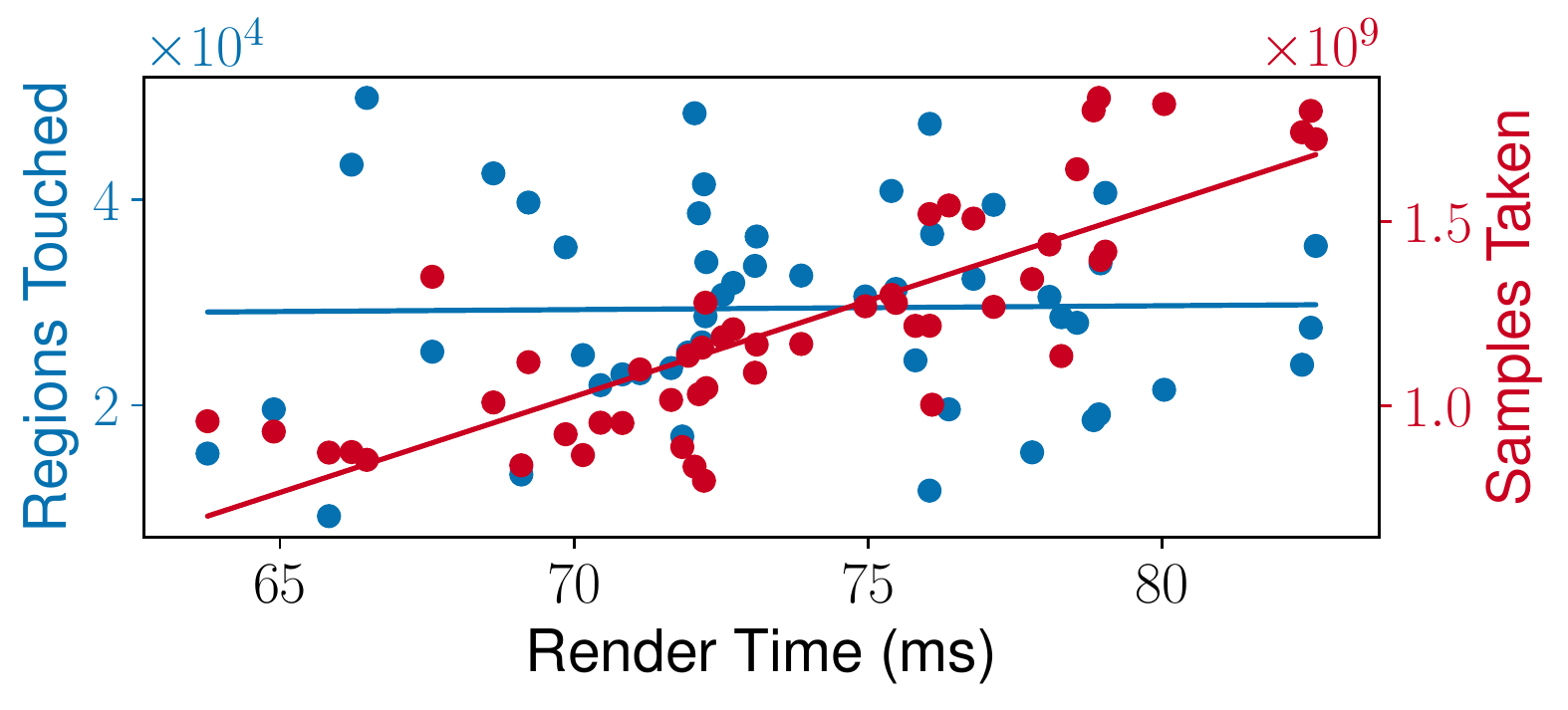}
        \vspace{-1.75em}
        \caption{\label{fig:landing_gear_surface_perf_v_regions}%
        NASA Landing Gear (iso-surface)}
    \end{subfigure}
    \begin{subfigure}{0.48\columnwidth}
        \centering
        \includegraphics[width=\textwidth]{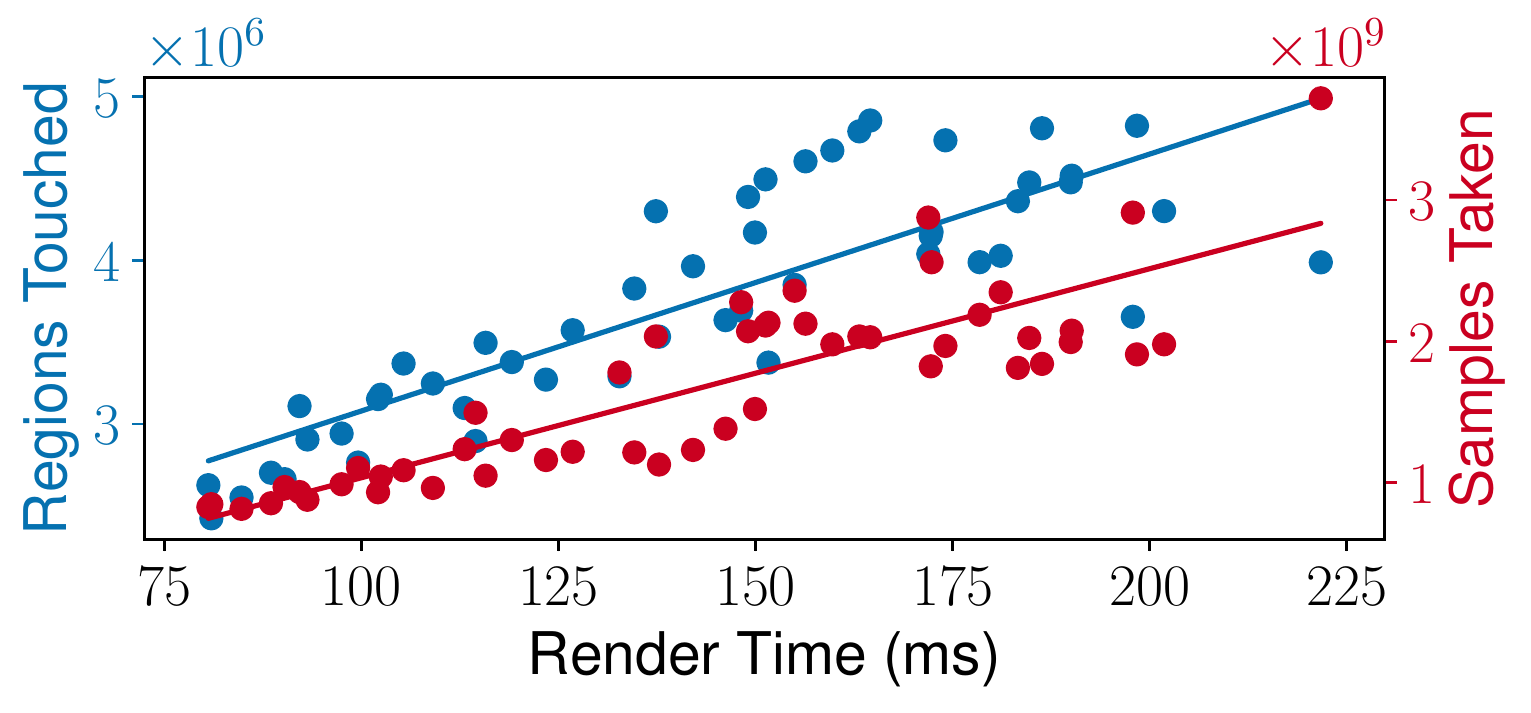}
        \vspace{-1.75em}
        \caption{\label{fig:exajet_rear_perf_v_regions}%
        Exajet (rear)}
    \end{subfigure}
    \begin{subfigure}{0.48\columnwidth}
        \centering
        \includegraphics[width=\textwidth]{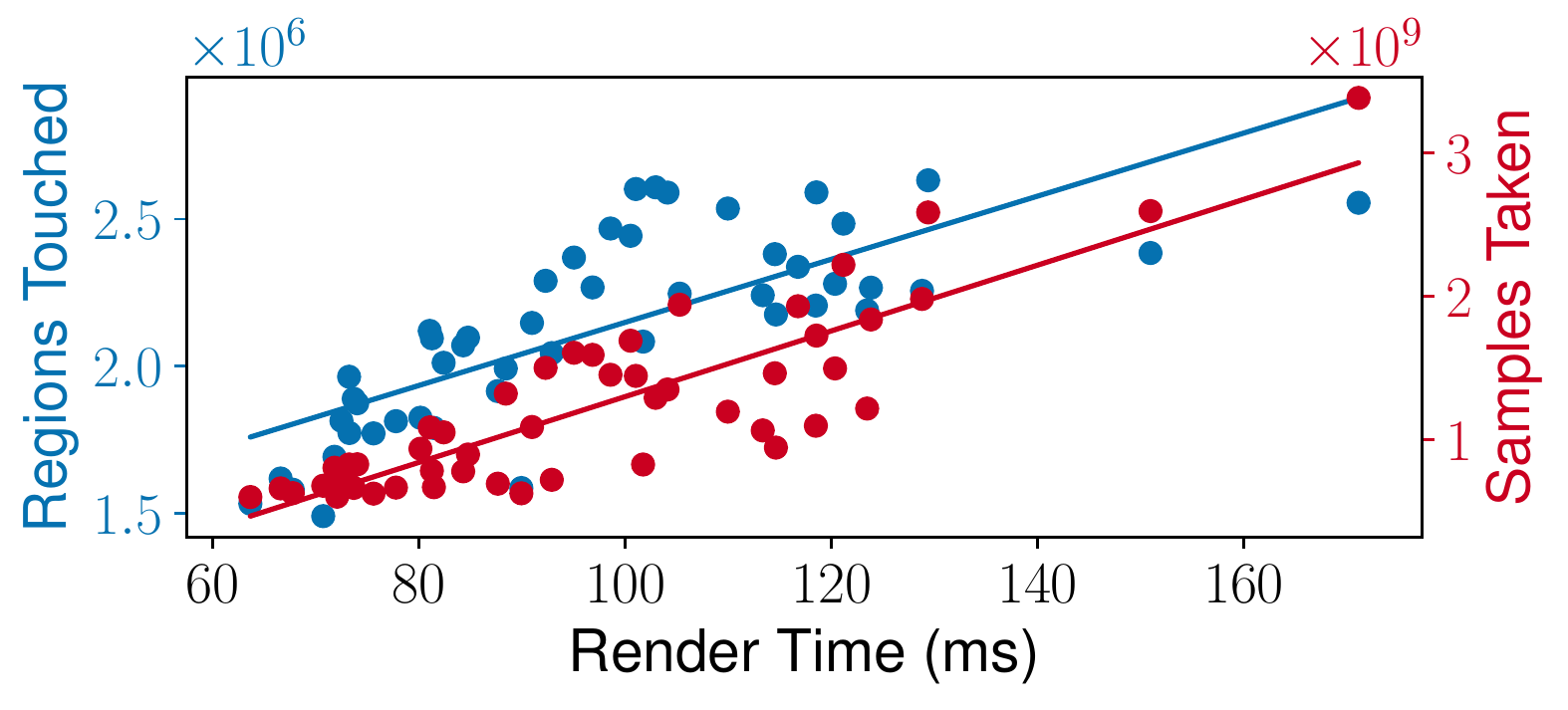}
        \vspace{-1.75em}
        \caption{\label{fig:exajet_wing_perf_v_regions}%
        Exajet (wing)}
    \end{subfigure}
    \vspace{-0.75em}
    \caption{\label{fig:perf_v_regions}%
    Rendering performance vs.\ the number of regions traversed and
    samples taken,
    measured over a 50 position spherical camera orbit.}
    \vspace{-2em}
\end{figure}

In \Cref{fig:teaser,fig:models} we report average rendering times
measured over 150 frames. These visualizations are representative of
typical use cases for performing high-quality visualizations of AMR
data; however, as with any volume renderer, the final performance is
strongly tied to the transfer function chosen, as this directly
affects space skipping, adaptive sample rates, and early ray
termination. Therefore we also performed a scalability study where we
moved the camera in a spherical orbit on the models' bounding spheres
to render 50 different viewpoints. We report
rendering performance in milliseconds, along with the number of
regions touched and samples taken, in \Cref{fig:perf_v_regions}.

While for the representative visualizations
(\Cref{fig:teaser,fig:models}) we terminate rays early
at an opacity threshold of 98\%, we disable early ray
termination for the orbit benchmark (\Cref{fig:perf_v_regions}),
to make the study less
dependent on occlusion from large, homogeneous features
(e.g., the ocean surface of the \impact). We observe an interesting
correlation between rendering performance and region size. Models that
have relatively few but large regions, e.g., the \stellar (170K
non-empty regions) or \gear (283K non-empty regions), require us to
take relatively more samples than models with smaller regions, e.g.,
the \cloud (2.54M non-empty regions) or \exajet rear (15.3M non-empty
regions). We also observe that performance on those models
is not correlated to the number of regions touched, which indicates
that the overhead of traversing the region BVH with RTX is
negligible compared to the cost for the large number of samples.
We also find that models with large regions with more cells have higher rendering times
than those with more cells but fewer cells per region, e.g., \exajet. This suggests
that models with more but smaller bricks can make better use of empty space
skipping to improve performance.

Across the benchmarks our method remains interactive, even
at the high quality settings chosen throughout the paper and the high
resolution used in \Cref{fig:teaser}. Moreover, if higher framerates
or higher resolutions are desired, the user could lower the
sampling rate to improve performance. We also note that, as an
image-parallel approach, our method scales well as more GPUs are added
to the system.

\subsection{Comparison to Existing Methods}

Apart from absolute performance numbers, 
adequately judging a method's performance and/or
quality is, generally speaking, more easily achieved by comparing it to state of the art techniques.
We note that such comparisons are notoriously hard to do,
as different frameworks support different hardware platforms,
rendering features, or illumination models.
We have identified two comparisons that stand out among the rest:
First, we compare the \emph{algorithmic} differences of our technique
against prior approaches proposed by K\"ahler and Abel~\cite{kaehler_single-pass_2013}
and Wald et al.~\cite{wald:17:AMR}
(\Cref{sec:results_kaehler_compare,sec:results_wald_compare}).
Second, we evaluate the performance of our complete framework against the most comparable
alternative framework for rendering large-scale AMR data,
OSPRay~\cite{ospray}, in \Cref{sec:results_ospray_compare}.




\subsubsection{Comparison to K\"ahler and Abel~\cite{kaehler_single-pass_2013}}
\label{sec:results_kaehler_compare}
As discussed previously, K\"ahler and Abel~\cite{kaehler_single-pass_2013} propose a similar
data structure to our own and construct their bricks and $k$-d tree
in a nearly identical manner. They also employ ray tracing and
adaptive sampling to render each brick.
The core difference between our method and theirs is that they only
target nearest neighbor reconstruction or vertex-centered
data, avoiding the key problem that our work addresses---fast and
high-quality rendering of cell-centered AMR data.
Arguably, if one were to
start with their framework, and incrementally add our features, such
as regions for fast basis reconstruction, iso-surfaces, and analytic
gradients, one would arrive at
exactly the methods and algorithms described in this paper.
Thus, one logical way of viewing our method is as building on
the same core ideas proposed by K\"ahler and Abel, and improving upon it by
adding a set of additional techniques such to support basis
reconstruction without cell lookups, iso-surfaces, analytic gradients,
and a more modern GPU implementation with RTX acceleration.


\ifx\empty
Restricting ourselves to only nearest reoncstruction, K\"ahler and Abel
use texture hardware for sampling in each brick, which we do
not. Partly this is because it is as yet unclear if or how texture
hardware can be used at all for basis-method reconstruction, partly it
is because for the kind of models we are targetting the majority of
bricks are far too small to be used with 3D hardware texturing (we
tried!). Finally, whereas K\"ahler and Abel use a $k$-d tree traversal to
trace a ray trhough its bricks, we use hardware ray tracing, which
back then was not available.
\fi

\subsubsection{Comparison to Wald et al.~\cite{wald:17:AMR}}
\label{sec:results_wald_compare}
A more recent method to compare against is that
of Wald et al.~\cite{wald:17:AMR}.
Wald et al. not only proposed the basis method used in
this work, but
a set of reconstruction kernels that have since been
further improved by Wang et al.~\cite{wang:20:tamr,wang:18:iso-amr}. Though it
\emph{should} be possible to add these kernels to our
framework we have not yet done so, and in particular can not yet
match the reconstruction quality of Wang et al.~\cite{wang:20:tamr}.

However, we can evaluate the impact of our data structure on
evaluating the basis method.
To do this, we modified our brick construction algorithm
to save the partitioning planes used during construction; producing
effectively the same $k$-d tree used by Wald et
al.~\cite{wald:17:AMR}. We then modified our
renderer to still use the regions to decide \emph{where} to sample,
but to ignore the list of active bricks and instead evaluate the basis method using
a software cell location
kernel similar to the one proposed in the original
paper~\cite{wald:17:AMR}.
Both methods are run using the same settings and thus take the same
samples. To keep the same rendering
framework we use our fast analytic gradients for both variants, even
though these were not available in the original method.
The results of this experiment are given in
Table~\ref{tab:cell-loc-vs-ours}. We find that our technique
leads to a speed-up of $1.9-5.1\times$ over the original basis
method, by eliminating additional cell location traversals.



\begin{table}
  \centering
  \caption{\label{tab:mode_comparisons}
    \label{tab:cell-loc-vs-ours}%
    Performance of reconstruction
    with per-sample cell location kernels in milliseconds as originally proposed by
    Wald et al.~\cite{wald:17:AMR}, vs. our reconstruction from the
    active brick regions (Section~\ref{sec:fast-basis-reconstruction}).
    \vspace{-1em}
    }
  \relsize{-1}{
    \begin{tabular}{@{}l|rrr@{}}
    \toprule
    Model
    &
    via cell loc.
    &
    from regions
    & speedup
    \\
    &
    (Wald et al.~\cite{wald:17:AMR})
    &
    (ours)
    \\
    \midrule
Cloud          & 88.5 & 37.8 & $2.3\times$\\
Impact-5K      & 80.6 & 38.9 & $2.1\times$\\
Impact-20K     & 258  & 130  & $2.0\times$\\
Impact-46K     & 391  & 202  & $1.9\times$\\
Wind           & 151  & 75.8 & $2.0\times$\\
Gear           & 288  & 56.5 & $5.1\times$\\ 
Exajet (rear)  & 218  & 80.6 & $2.7\times$\\ 
Exajet (wing)  & 88.5 & 36.2 & $2.4\times$\\
    \bottomrule
    \end{tabular}
  }
\end{table}

\subsubsection{Comparison to OSPRay}
\label{sec:results_ospray_compare}
\begin{figure}[b]
  \vspace{-1.5em}
  \centering
  \setlength{\tabcolsep}{.1ex}
{\relsize{-1}{ 
    \textsf{
    \begin{tabular}{@{}ccc@{}}
        \toprule
    Ours & \multicolumn{2}{c}{OSPRay AMR} \\
    (Default) & Similar Quality & Similar Perf.\\
    \midrule
    \includegraphics[width=.32\columnwidth]{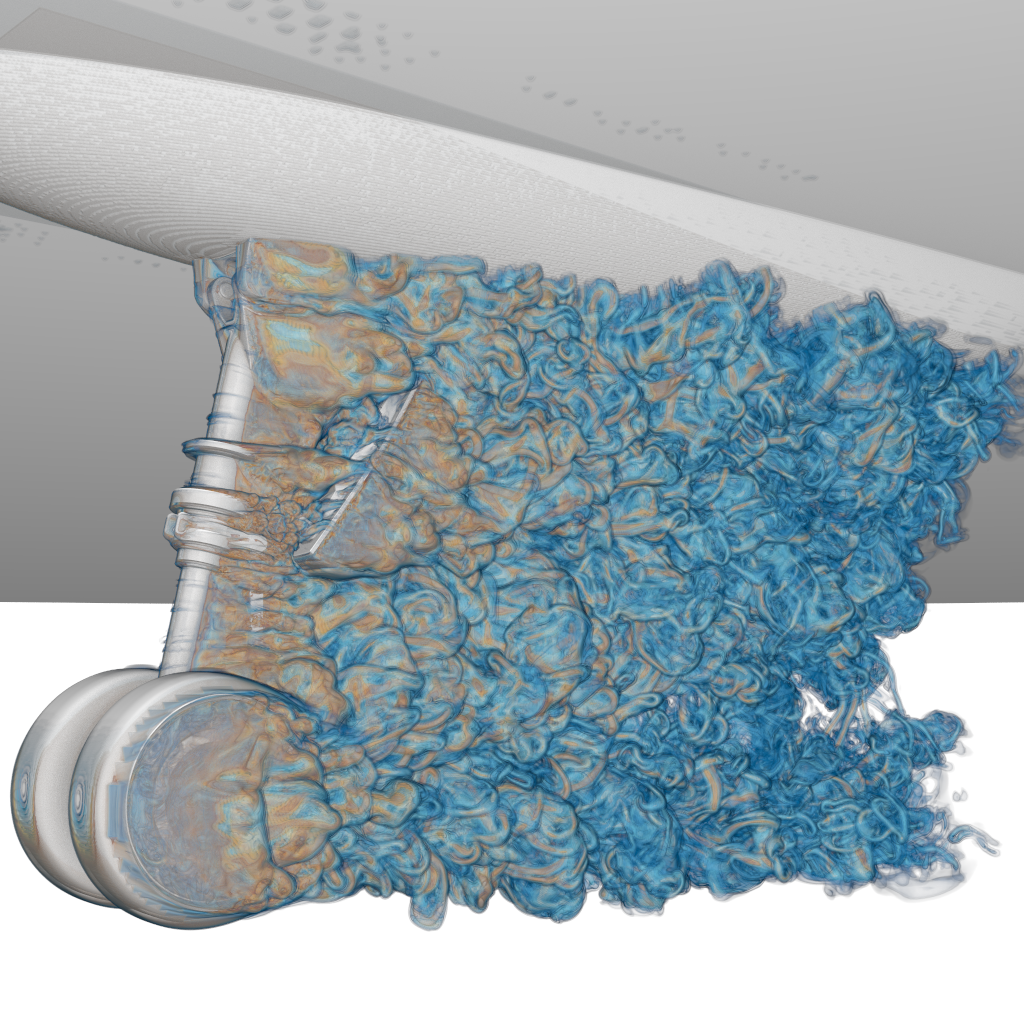}
    &
    \includegraphics[width=.32\columnwidth]{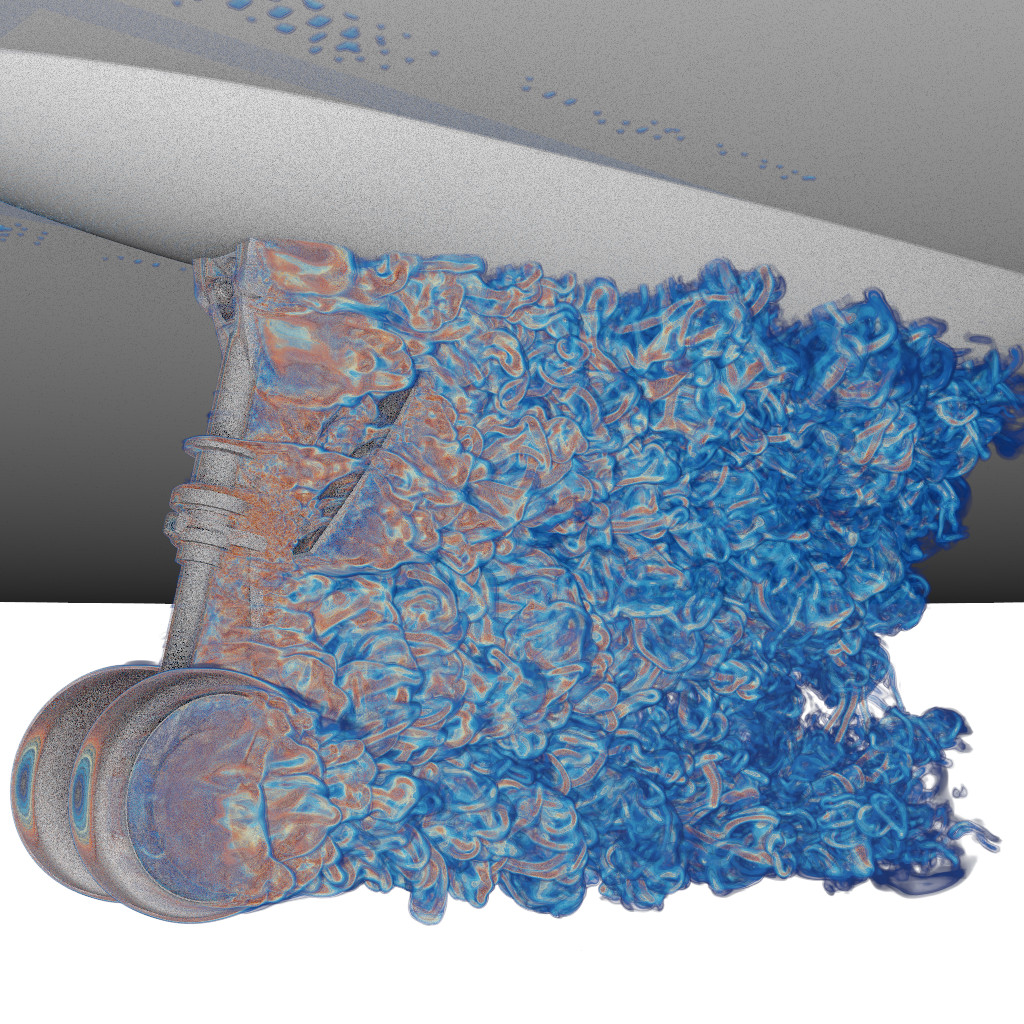}
    &
    \includegraphics[width=.32\columnwidth]{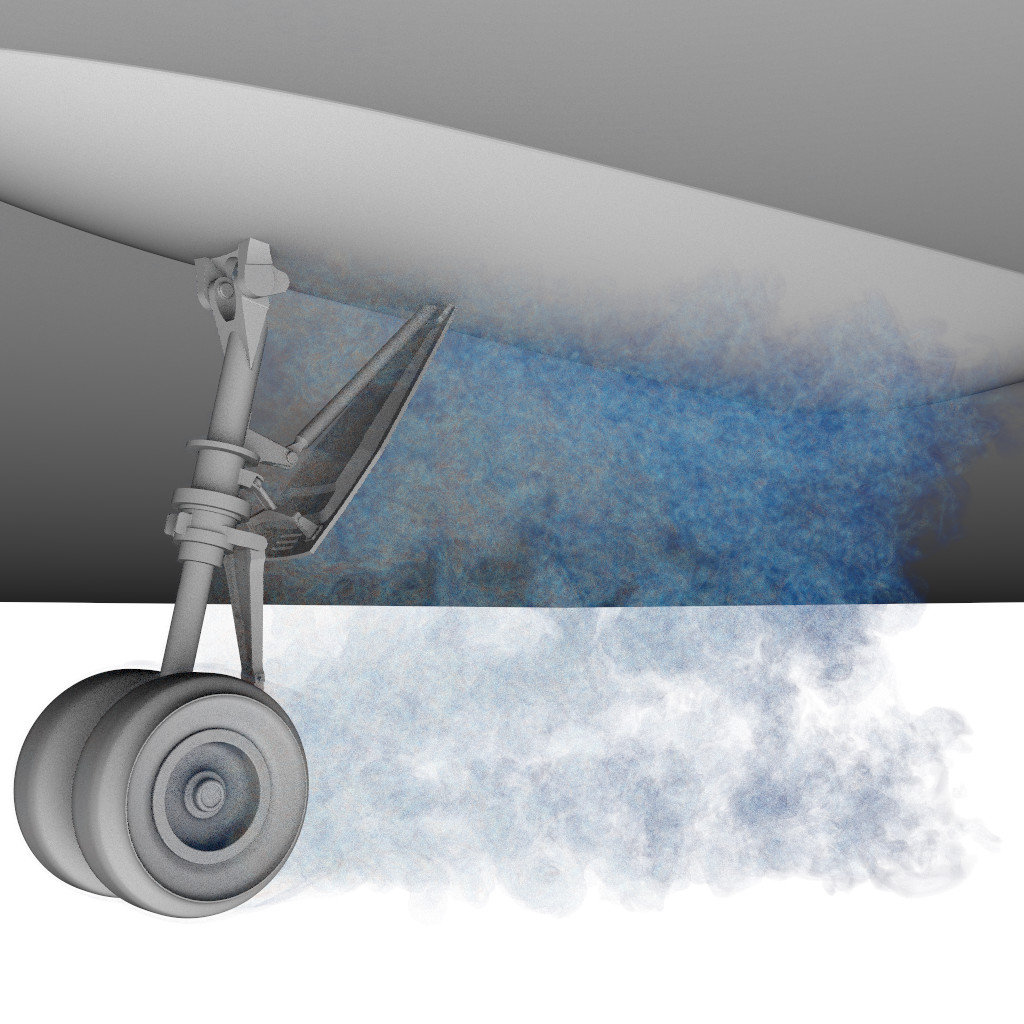}
    \\
    (a)~0.06~sec & (b)~12.5~sec & (c)~0.24~sec\\
    \bottomrule
  \end{tabular}
  }
  }
  }
    \vspace{-0.5em}
    \caption{\label{fig:lg_ospray_amr_comparison}\label{fig:ours-vs-ospray-amr}%
      Comparison to
      OSPRay's native AMR rendering on the \gear
      (b) targeting similar quality or (c) performance
      OSPRay's current AMR traversal does not leverage the
      underlying hierarchy for space skipping or adaptive
      sampling, leading to either poor performance at good quality (b), or
      poor quality for interactive performance (c).}
\end{figure}

\begin{figure}
  \centering
  \setlength{\tabcolsep}{.1ex}
  {\relsize{-1}{
      \textsf{
      \begin{tabular}{@{}ccc@{}}
      \toprule
          Ours & \multicolumn{2}{c}{OSPRay Unstructured} \\
          (Default) & Similar Quality & Similar Perf.\\
    \midrule
        \includegraphics[width=.32\columnwidth]{meteor-20k-cut.png}
    &
        \includegraphics[width=.32\columnwidth]{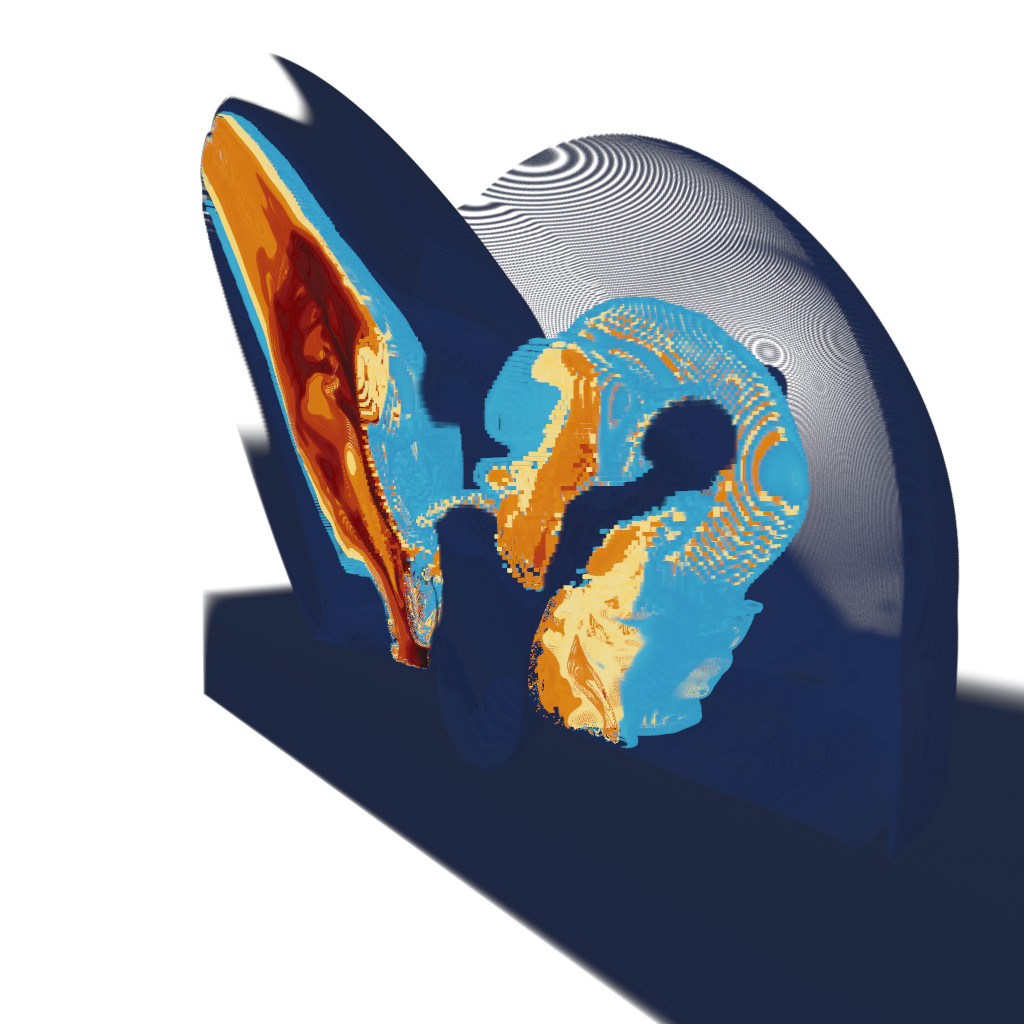}
    &
        \includegraphics[width=.32\columnwidth]{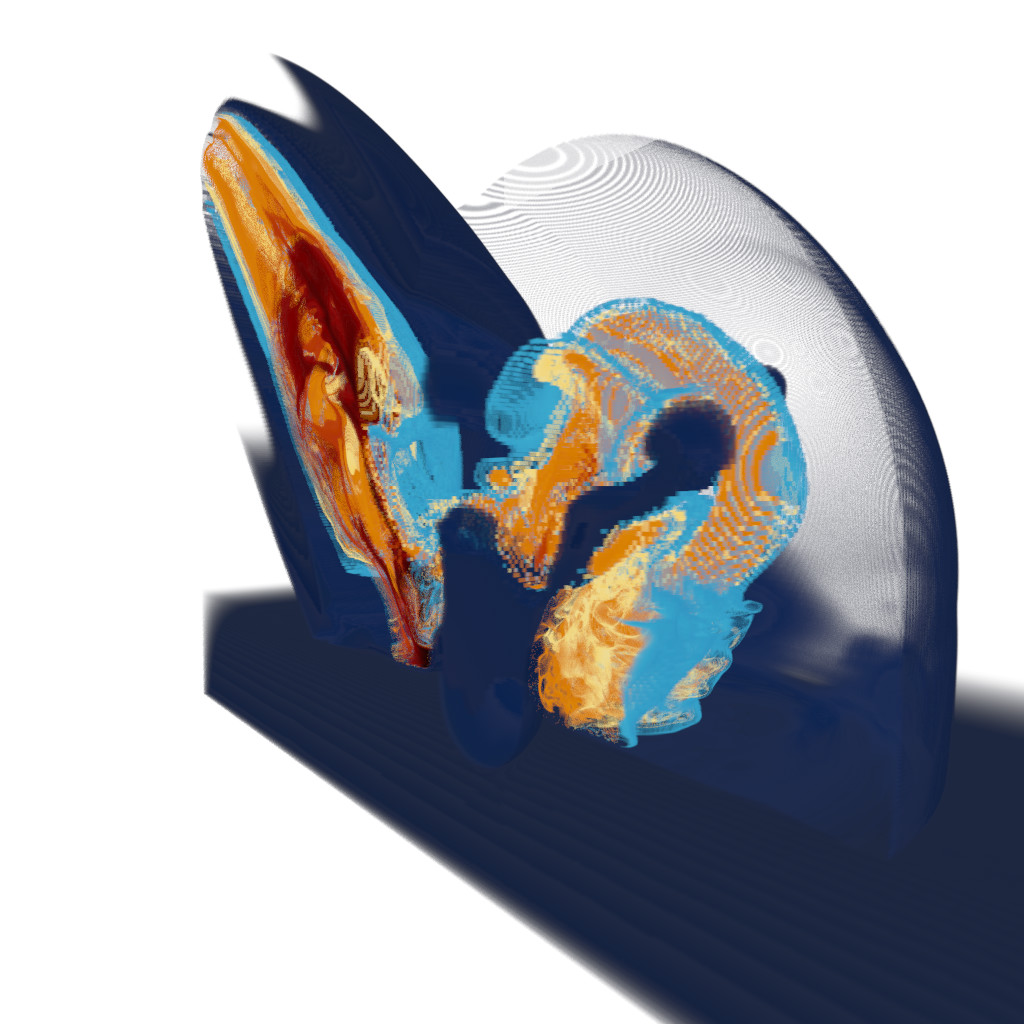}
        \\
        (a) 69.6~ms & (b) 435~ms & (c) 133~ms\\
        \includegraphics[width=.32\columnwidth]{exajet-vel-mag-rear.png}
    &
        \includegraphics[width=.32\columnwidth]{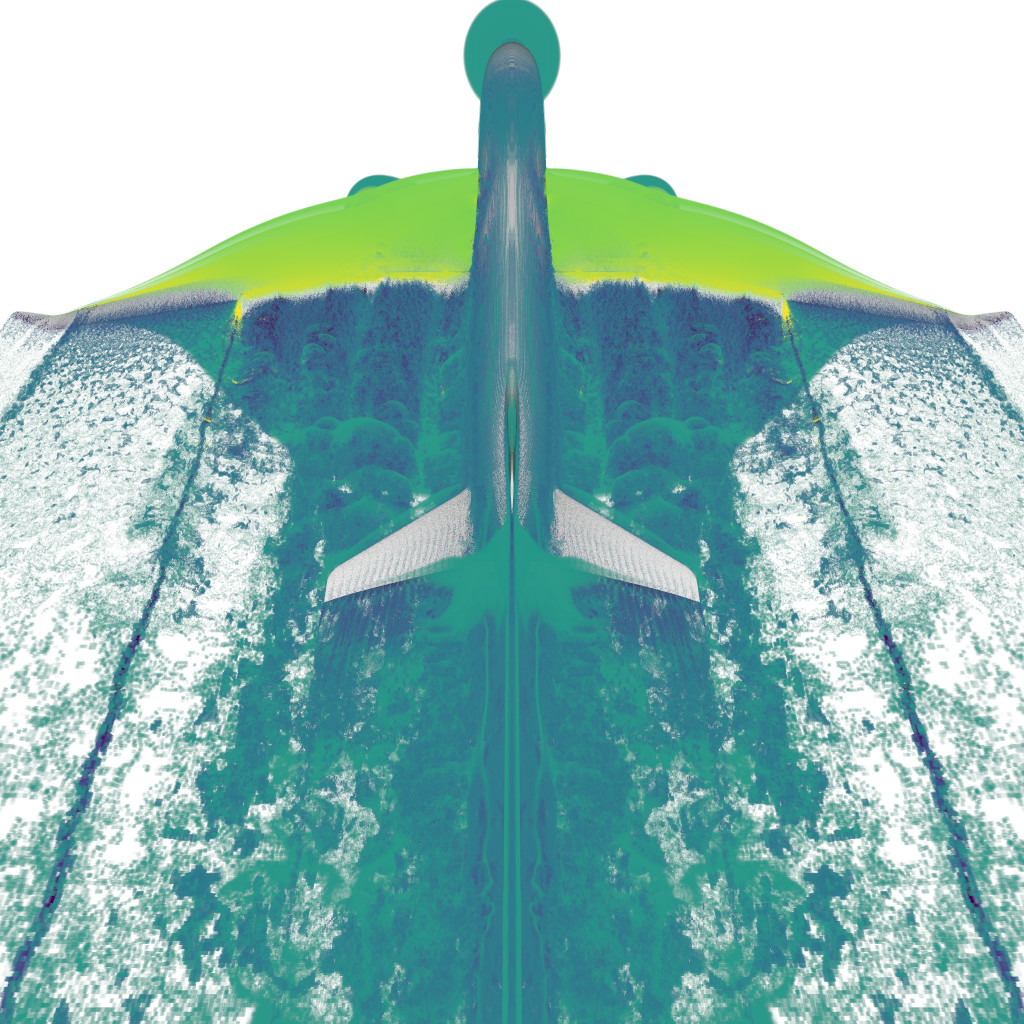}
    &
        \includegraphics[width=.32\columnwidth]{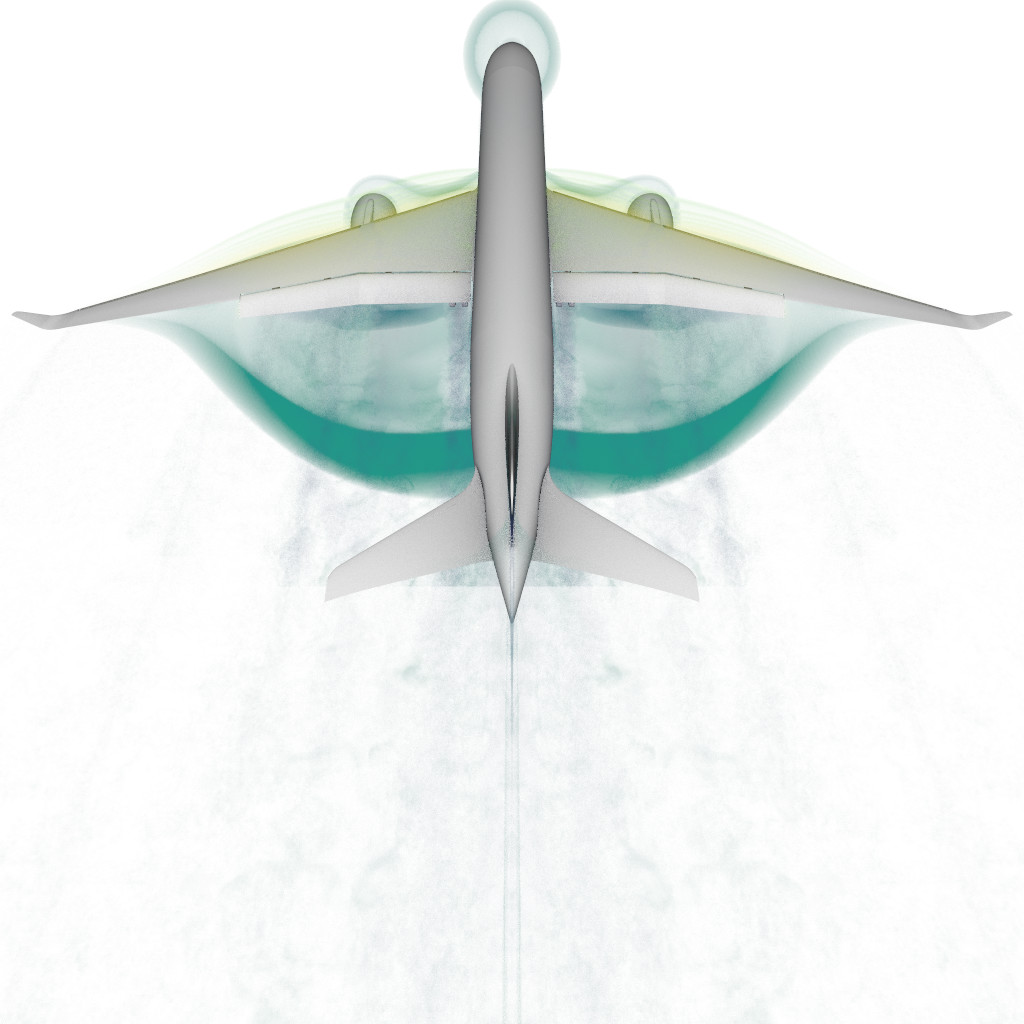}
        \\
        (d) 80.1~ms & (e) 3.33~sec & (f) 185~ms\\
        \bottomrule
  \end{tabular}
  }
  }
    }
    \vspace{-0.5em}
    \caption{\label{fig:ospray_quality_perf_unstructured}%
    Quality and performance comparisons against rendering the AMR data
    as an unstructured mesh using OSPRay.
    The unstructured mesh requires significantly more memory (b,c: 40~GB,
    e,f: 160~GB), and trading performance for image quality (b,e)
    or vice-versa (c, f).
    \vspace{-2em}}
\end{figure}

To compare our approach to the state of the art volume rendering
supported in OSPRay we took the latest version of OSPRay,
imported our data sets, and tuned the transfer function and
sampling rate to achieve either similar performance or roughly similar
quality.
We note that comparisons between two different frameworks are
always apples-to-oranges, and that the output images will never
fully match. For example, our method performs adaptive sampling,
while OSPRay ray marches at a fixed step size; similarly, our method
always performs gradient shading and lighting, while OSPRay does not
(compare, e.g., \Cref{fig:ospray_quality_perf_unstructured}a and b).
For these experiments, we
ran OSPRay on a high-end workstation equipped with a Xeon W-3275M
28-core ``Cascade Lake'' CPU and 256~GB of RAM.



OSPRay provides native support for rendering Block-Structured AMR,
such as the \gear, which we perform a direct comparison with in
\Cref{fig:ours-vs-ospray-amr}.
Though any such comparison has to be taken with a grain of salt, the general
observation is that our system achieves either significantly higher
quality at similar performance,
or higher performance at similar quality.
Although OSPRay offers native support for Block-Structured AMR,
it does not support Tree-Based AMR variants,
a limitation shared by standard tools such as ParaView and VisIt.
For AMR data that does not fit OSPRay's requirements (e.g., \exajet
and \impact), an alternative approach to rendering such data
is to first convert it to an unstructured mesh by computing
its dual mesh, then render the resulting mesh using OSPRay's
unstructured mesh renderer.


While flattening the data allows scientists to visualize it using
standard tools, it is far from ideal. The resulting unstructured
mesh occupies significantly more memory than the original data and is
more challenging to render due to the now unstructured layout
(\Cref{fig:ospray_quality_perf_unstructured}). Consequently the
observation of ``better performance at similar quality or better quality
at similar performance'' is more pronounced in this comparison
(\Cref{fig:ospray_quality_perf_unstructured}).

\section{Limitations}
\label{sec:discussion}
\label{sec:limitations}

Our presented ExaBricks data structure is suitable for high-quality
interactive rendering of the large scale AMR data sets used in
simulations today.
However, our approach is not without limitations.
We currently only support the \emph{basis} method for
sample reconstruction, and while methods such as
\emph{octant}~\cite{wang:18:iso-amr} or \emph{GTI}~\cite{wang:20:tamr}
should be possible to add, doing so would require
reexamining what the support of a cell (and thus, brick) looks like.


Another potential limitation of our method is that its performance and
memory consumption depend on the number and distribution of
bricks a data set is partitioned into. For example, on the \exajet
the curved empty regions around the plane geometry result in a large
number of single-cell bricks. While our space skipping and adaptive
sampling strategies reduce the impact of such bricks, they do not
come without cost. A potential option to address this would be to allow
for partially filled bricks.
Similarly, the number and distribution of the active brick regions can
become non-trivial, challenging the BVH construction or
traversal. To this end, future work on improving the builder to produce less,
and possibly less exact, regions could be valuable.
Finally, we do not take advantage of hardware texture interpolation
in our method, as the large number of bricks may require a
substantial number of textures or a large atlas. However, this could further improve
rendering performance by accelerating sample computation.





Arguably more important than the output brick or region distribution,
is that neither is computed in real time in our current implementation.
The bricks are constructed in an offline
pre-process and can take several minutes, while the regions are built
at load time and can take tens of seconds.
Both steps can likely be accelerated significantly, and would
improve the experience for future end users.
From a practical standpoint, the biggest limitation of our framework
is that it is not yet available within a standard
visualization package. Although we plan to make our source code
available, making these capabilities available to end users would
require an integration into ParaView~\cite{paraview} or VisIt~\cite{visit}.
Such an integration would address a
clear need in the AMR community and could be done through an
integration into OSPRay~\cite{ospray} or IndeX~\cite{nvidia_index}, which are
already used by these tools.



\section{Conclusion}
\label{sec:conclusion}

In this paper, we have proposed a novel approach for high-quality
and efficient rendering of AMR data, through a combination of three
inter-operating data structures. We have demonstrated our method's
capabilities on a set of non-trivial models, and have shown that 
it is capable of achieving interactive
performance on a single workstation for large models and
demanding high-quality rendering settings,
and is competitive with the state of the art.
A key advantage of our method is its generality, both in how easily
it can be adapted to support different AMR formats, and in how
it lends itself to implementation in other rendering frameworks.
Although the presented implementation leverages GPU hardware ray
tracing, there is no reason the same approach could not be
implemented in a CPU framework such as OSPRay.


The key to our approach is our ExaBricks data structure---and in
particular the Active Brick Regions---which provide
the ``glue'' that allows for seamlessly combining several known
techniques (bricking AMR data, \emph{basis} reconstruction,
space skipping, adaptive sampling, RTX volume traversal)
into a single algorithmic
framework in which they operate on the same
set of active brick regions and work to each other's advantage.
Our approach for combining these techniques
forms a compelling blueprint for high-quality AMR rendering,
and serves as an example which may be adopted in standard
visualization packages.



\label{sec:conclusion}



\acknowledgments{
    The Landing Gear was graciously provided by Michael Barad, Cetin
    Kiris and Pat Moran of NASA. The Exajet was made available
    by Exa GmbH and Pat Moran.
    The TAC Molecular Cloud is courtesy of Daniel Seifried.
    The Stellar Cluster Wind is courtesy of Melinda
    Soares-Furtado.
    This work was supported in part by
    NSF OAC awards 1842042, 1941085,
    NSF CMMI award 1629660,
    LLNL LDRD project SI-20-001
    This material is based in part upon work supported by the
    Department of Energy (DoE), National Nuclear Security Administration
    (NNSA), under award DE-NA0002375.
    This research was supported in part by the Exascale Computing Project (17-SC-20-SC),
    a collaborative effort of the DoE Office of Science and the NNSA.
    This work was performed in part under the auspices of the DoE by Lawrence Livermore National
    Laboratory under Contract DE-AC52-07NA27344.}
\bibliographystyle{abbrv-doi}
\bibliography{exabrick,will}






\end{document}